\documentclass[a4paper]{article}
\pdfoutput=1
\usepackage{jhepmod}

\usepackage{multirow}
\usepackage{hyperref}
\usepackage{graphicx}
\usepackage{mathtools}
\usepackage{amsmath,amssymb,amsfonts,amsthm,latexsym,stmaryrd,physics,mathrsfs}
\allowdisplaybreaks[4]
\usepackage{marginnote}
\usepackage{color}
\usepackage{bm}
\usepackage{float}
\usepackage{nicefrac}
\usepackage{microtype} 
\usepackage{booktabs}
\usepackage{verbatim}
\usepackage{setspace}
\usepackage[utf8]{inputenc}
\usepackage{stfloats}
\usepackage{diagbox}
\usepackage{dsfont}
\usepackage{longtable}
\usepackage{afterpage}
\usepackage{bbold}
\usepackage{multirow}
\usepackage{multirow}
\usepackage{ulem}
\usepackage[colorinlistoftodos,prependcaption,textsize=tiny]{todonotes}

\def\beq{\begin{equation}}
\def\eeq{\end{equation}}
\newcommand{\bea}{\begin{eqnarray}}
\newcommand{\eea}{\end{eqnarray}}
\def\bi{\begin{itemize}}
\def\ei{\end{itemize}}
\def\ba{\begin{array}}
\def\ea{\end{array}}
\def\bfig{\begin{figure}}
\def\efig{\end{figure}}

\def\C{\mathbb{C}}
\def\R{\mathbb{R}}
\def\Z{\mathbb{Z}}

\def\sgn{\text{sgn}}

\newcommand{\Slc}{\mathrm{SL}(2,\mathbb{C})}

\newcommand{\Su}{\mathrm{SU}(2)}
\newcommand{\Suone}{\mathrm{SU}(1,1)}

\def\be{\begin{eqnarray}}
\def\ee{\end{eqnarray}}

\newcommand{\ck}{\mathcal K}

\newcommand{\cn}{\mathcal N}

\newcommand{\cs}{\mathcal S}

\newcommand{\fg}{\mathfrak{g}}

\newcommand{\fp}{\mathfrak{p}}

  \newcommand{\Fs}{\mathfrak{S}}

\renewcommand{\a}{\alpha}
\renewcommand{\b}{\beta}
\newcommand{\g}{\gamma}

\newcommand{\sig}{\sigma}

\renewcommand{\l}{\lambda}
\renewcommand{\L }{\Lambda}

\newcommand{\rmd}{\mathrm d}

\newcommand{\lt}{\left}
\newcommand{\rt}{\right}

\newcommand{\re}{\mathrm{Re}}

\title{A Mathematica program for numerically computing real and complex critical points in 4-dimensional Lorentzian spinfoam amplitude}

\author[1,2]{Muxin Han}
\author[2]{\ Hongguang Liu}
\author[3]{\ Dongxue Qu}
\affiliation[1]{Department of Physics, Florida Atlantic University, 777 Glades Road, Boca Raton, FL 33431-0991, USA}
\affiliation[2]{Department Physik, Institut f\"ur Quantengravitation, Theoretische Physik III, Friedrich-Alexander Universit\"at Erlangen-N\"urnberg, Staudtstr. 7/B2, 91058 Erlangen, Germany}
\affiliation[3]{Perimeter Institute for Theoretical Physics, 31 Caroline St N, Waterloo, ON N2L 2Y5, Canada}

\emailAdd{hanm(AT)fau.edu}
\emailAdd{hongguang.liu(AT)gravity.fau.de}
\emailAdd{dqu(AT)perimeterinstitute.ca}

\abstract{This work develops a comprehensive algorithm and a Mathematica program to construct boundary data and compute real and complex critical points in spinfoam amplitudes. Our approach covers both spacelike tetrahedra and triangles in the EPRL model and timelike tetrahedra and triangles in the Conrady-Hnybida extension, aiming at addressing a wide range of physical scenarios such as cosmology and black holes. Starting with a single 4-simplex, we explain how to numerically construct boundary data and corresponding real critical points from any nondegenerate 4-simplex geometry. Extending this to the simplicial complex, we demonstrate the algorithm for constructing boundary data and critical points using examples with two 4-simplices sharing an internal tetrahedron. By revisiting the $\Delta_3$ triangulation with curved geometry, we demonstrate the numerical computation of the real critical point corresponding to the flat geometry and the deformation to the complex critical points. Additionally, the program evaluates the spinfoam action at the critical points and compare to the Regge action. }

\begin{document}

\maketitle

\section{Introduction}
In recent years, there has been growing attention devoted to exploring real and complex critical points in 4-dimensional Lorentzian spinfoam amplitudes. Critical points play a crucial role in the stationary phase approximation, as a powerful technique for studying quantum theory by perturbative expansion. Studying real and complex critical points in the spinfoam amplitude closely relates to recent progress in numerical analysis of spinfoam amplitudes \cite{Dona:2022yyn, Han:2021kll}. The recent results relate the critical points and stationary phase approximation to the effective dynamics from the spinfoam amplitude, which can be applied to physical scenarios such as quantum cosmology and black holes \cite{Han:2023cen, Han:2024ydv, Han:2024rqb}. Given the complexity of the spinfoam amplitude, the critical point and its corresponding contribution to the spinfoam amplitude have to be computed numerically. Another related numerical result is the semiclassical expansion of the spinfoam amplitude to the next-to-leading order from the stationary phase approximation \cite{Han:2020fil, toappear15}. We also would like to mention a few other numerical approaches for spinfoam quantum gravity, including the ``sl2cfoam-next" program for the non-perturbative computation of the spinfoam amplitude \cite{Gozzini:2021kbt, Frisoni:2022urv, Dona:2022dxs}, the effective spinfoam model \cite{Asante:2020qpa, Asante:2021zzh}, the hybrid algorithm \cite{Asante:2022lnp}, and spinfoam renormalization \cite{Asante:2022dnj, Bahr:2016hwc}, etc. 

The early stationary phase analysis of spinfoam amplitudes has primarily focused on understanding how to classify critical points mathematically and how discrete geometries can be reconstructed from real critical points (see e.g. \cite{Barrett:2009mw,Han:2011re}). Although these results are important from the conceptual understanding of spinfoams, there is a growing recognition of the necessity to shift perspectives: The reverse procedure, i.e. constructing the real critical point of spinfoam amplitudes from discrete geometry, turns out to be more practical in the computation of spinfoam amplitude. The reason is that constructing a real critical point is the starting point of the stationary phase approximation, and even if we need a complex critical point, it has to be a deformation from a real critical point. This change in perspective inspires the development of computational tools capable of efficiently identifying and computing real and complex critical points in spinfoam amplitudes. This motivation drives us to develop a comprehensive Mathematica program specifically designed to identify real and complex critical points based on geometric interpretations. This program not only facilitates the computation process, but also aims to help researchers understand the relation between discrete geometries and their associated critical points in the 4-dimensional Lorentzian spinfoam formalism. 

In this work, we present the algorithm and corresponding program for constructing boundary data and computing real and complex critical points based on the nondegenerate Lorentzian geometry of a simplicial complex. The relevant code, along with Mathematica notebooks, is provided in \cite{numericalResult}. Our algorithm and program cover both the Lorentzian Engle-Pereira-Rovelli-Livine (EPRL) model \cite{Engle:2007wy, Livine:2024hhc} with only spacelike tetrahedra and triangles and the Conrady-Hnybida extension \cite{Conrady:2010kc, Conrady:2010vx} that includes timelike tetrahedra and triangles. By addressing both spacelike and timelike objects, we aim to cover a broad range of scenarios commonly encountered in physical applications (e.g. the application to cosmology \cite{Han:2024ydv}), thereby improving the practicality of our algorithm. Our present work develops a Mathematica program to perform the computation, whereas the algorithm may be implemented in any other computer language.

This paper is organized as follows: Section \ref{SFAmplitude} gives a brief review of the integral representation of the spinfoam amplitude and the definition of the large-$j$ regime. In Section \ref{v1data}, we explain how to construct the boundary data and compute the real critical points on a 4-simplex with both spacelike and timelike triangles. We also introduce the parity transform on a 4-simplex. Section \ref{SimplicialComp} generalizes the computation to the real critical point of the spinfoam amplitude on a simplicial complex and demonstrates the algorithm. In this section, we consider an example of a complex containing two 4-simplices sharing a common internal tetrahedron. We discuss two cases: one with the spacelike internal tetrahedron and the other with timelike internal tetrahedron. Section \ref{CCP} revisits the known results on the spinfoam amplitude on the $\Delta_3$ complex as an example to demonstrate the algorithm of computing the real and complex critical points in spinfoams. In Section \ref{Conclusion and Discussion}, we conclude and discuss some outlooks.

\section{Spinfoam amplitude}\label{SFAmplitude}

A 4-dimensional simplicial complex $\ck$ contains 4-simplices $v$, tetrahedra $e$, triangles $f$, edges, and vertices. The internal and boundary triangles are denoted by $h$ and $b$, and $f$ denotes a triangle which is either $h$ or $b$. In this paper, we consider the spinfoam amplitude on $\ck$ with the Conrady-Hnybida extension, which allows not only spacelike tetrahedra and triangles but also timelike tetrahedra and triangles \cite{Conrady:2010vx,Conrady:2010kc}. The half-integer spins $j_h,j_b\in \mathbb{N}_0/2$, assigned to internal and boundary triangles $h,b$, label the irreducible representations of SU(2) and SU(1,1) for triangles belonging to spacelike and timelike tetrahedra,  respectively. The spins are the quanta of triangle areas. In the large-$j$ regime, the area of triangle $f$ is given by $\mathrm{Ar}_f\simeq \g j_f$ for spacelike triangles \cite{Rovelli1995,ALarea} and $\mathrm{Ar}_f=j_f$ for timelike triangles \footnote{For timelike triangles, $j_f$ here is defined as $ j_f \equiv \frac{n_f}{2}, n_f \in \mathbb{Z}_{+}$ where $(\rho_f,n_f)$ is the label of $\text{SL}(2,\mathbb{C})$ unitary irreducible representations used in \cite{Conrady:2010vx,Liu:2018gfc}}
, when we set the unit such that $8\pi G\hbar=1$. $\g$ is the Barbero-Immirzi parameter.

The Lorentzian spinfoam amplitude on $\ck$ sums over internal spins $\{j_h\}$:
\be
A(\mathcal{K})=\sum_{\{j_h\}}^{j^{\rm max}} \prod_h \mu_h(j_h) \int[\mathrm{d} X] e^{S[j_h, X;j_b,\xi_{eb}]}. \label{SFamplitude}
\ee  
We assume all boundary tetrahedra and triangles are spacelike, except for the discussion of 4-simplex amplitude in Section \ref{v1data}. The boundary states of $A(\ck)$ are SU(2) coherent states $|j_b,\xi_{eb}\rangle$, where $\xi_{eb}=u_{eb} (1,0)^\mathrm{T}$, and $u_{eb}\in \Su$. $j_b$ and $\xi_{eb}$ are determined by the area and the 3-normal of the triangle $b$ in the boundary tetrahedron $e$. $\mu_h(j_h)$ is the face amplitude, which we do not specify in this paper and does not affect our discussion\footnote{For the EPRL amplitude where all tetrahedra and triangles are spacelike, there is an argument to fix $\mu_h=(2j_h+1)^{1+|V(h)|}$, where $|V(h)|$ is the number of 4-simplices sharing $h$ \cite{Bianchi:2010fj,Han:2013gna}. But there has not yet been any argument to fix the face amplitude in the Conrady-Hnybida extension in the literature.}. The cut-offs of the spin sums, denoted by $j^{\rm max}=\{j^{\rm max}_h\}_h$, may be implied by the triangle inequality and $j_b$ or otherwise have to be imposed by hand in order to regularize the infrared divergence of the amplitude.  

The set of integrated variables, denoted by $X$, includes some $\Slc$ group elements $g_{ve}$ and some spinor variables. The spinfoam action $S$ in (\ref{SFamplitude}) is complex and linear with respect to $j_h$ and $j_b$. The action $S=\sum_f S_f$ is a sum of face actions $S_f$ and has three types of contributions from (1) spacelike triangles in spacelike tetrahedra, (2) spacelike triangles in timelike tetrahedra, and (3) timelike triangles.    Some details of the spinfoam action and integration variables are discussed below. Further details regarding the derivation of the spinfoam action $S$ can be found in \cite{Han:2013gna,Liu:2018gfc,Han:2021rjo}.

\subsection{Half-edge action}

Various types of triangles are related to different contributions to $S$, and they are associated with various types of variables in $X$, resulting in different kinds of contributions to $S$. A building block of $S$ is the 'half-edge action' $S_{vef}$. Here, $(v,e)$ corresponds to a half-edge in the dual complex $\mathcal{K}^*$. There are two types of half-edge actions corresponding to spacelike and timelike triangle $f$, respectively.

\vspace{2mm}
\textit{Spacelike triangle $f$:} The half-edge action $S_{vef}$ of spacelike triangles depend on the following variables: the spin $j_f\in\mathbb{N}_0/2$, the group variable $\fg_{ve}\in\Slc$, the spinor $z_{vf}\in\mathbb{CP}^1$, and the spinor $\tilde{\xi}_{ef}\in\C^2$ that is either SU(2) or SU(1,1). $\tilde{\xi}_{ef}\equiv\xi_{ef}$ is an SU(2) spinor if the tetrahedron $e$ is spacelike, while $\tilde{\xi}_{ef}\equiv\xi_{ef}^+$ or $\xi_{ef}^-$ is an SU(1,1) spinor if $e$ is timelike. The SU(2) and SU(1,1) spinors are given respectively by $\xi_{ef}=u_{ef}\xi_0^+$ and $\xi_{ef}^{\pm}=v_{ef}\xi_{0}^\pm \in \mathbb{C}^2$, where $u_{ef}\in\mathrm{SU(2)} $, $v_{ef}\in\mathrm{SU}(1,1)$ and  $\xi_0^+ = (1,0)^T$, $\xi_0^- = (0, 1)^T$. The spinor $z_{vf}$ is shared by different tetrahedra $e$ within the 4-simplex $v$. We define $Z_{vef}=\fg_{ve}^Tz_{vf}$.

The expression of the half-edge action $S_{vef}$ is given by
\be
S_{vef}&=&j_f\Bigg\{2\ln\left[\Bigl(m_{ef} \langle\tilde{\xi}_{ef}, Z_{vef}\rangle\Bigr)^{\frac{\kappa_{vef}+\det\eta_{e}}{2}}\Bigl(m_{ef} \langle Z_{vef},\tilde{\xi}_{ef}\rangle\Bigr)^{\frac{-\kappa_{vef}+\det\eta_{e}}{2}}\right]\nonumber\\
&&\qquad + \left(\imath\gamma \kappa_{vef}-\det\eta_e\right)\ln\Bigl[m_{ef}\langle Z_{vef},Z_{vef}\rangle\Bigr]\Bigg\}.\label{spacef}
\ee
$\det \eta_e=1$ (or $-1$) for spacelike (or timelike) tetrahedron $e$. We denote $v^*,e^*,f^*$ in the dual complex $\ck^*$ as the dual of $v,e,f$. The orientation of the face $f^*$ determines the orientation of $\partial f^*$. $\kappa_{vef}=\pm 1$ denotes the orientation of $f^*$ coinciding with ($\kappa_{vef}=+1$) or opposite to ($\kappa_{vef}=-1$) the orientation of the dual half-edge $(v, e)$, assuming the half-edge is always oriented outgoing from $v^*$. $\kappa_{vef}$ satisfies $\kappa_{vef}=-\kappa_{v'ef}=-\kappa_{ve'f}$. Furthermore, When $e$ is a spacelike, $\langle a,b\rangle=a^\dagger b$ is the SU(2) invariant inner product. When $e$ is timelike, $\langle a,b\rangle=a^\dagger\sig^3 b$ is the SU(1,1) invariant inner product. $m_{ef}\equiv\langle \xi_{ef}^{\pm},\xi_{ef}^{\pm}\rangle=\pm 1$, and the integration in \eqref{SFamplitude} is restricted to the domain where $m_{ef}\langle Z_{vef}, Z_{vef} \rangle>0$. 

\vspace{2mm}

\textit{Timelike triangle $f$:}  The half-edge action $S_{vef}$ of timelike triangles depends on the following variables in \eqref{SFamplitude}: the spin $j_f\in\mathbb{N}_0/2$, the group variable $\fg_{ve}\in\Slc$, the spinor $z_{vf}\in\mathbb{CP}^1$, and the SU(1,1) spinor $l_{ef}^+\in\C^2$. Here we define $l_{ef}^{\pm}=v_{ef} l_0^\pm \in \mathbb{C}^2$ where $v_{ef}\in \Suone$ and $l_0^{\pm} = (1,\pm1)$.

The half-edge action associated with the timelike triangle has four types: $S^+$, $S^-$, $S^{x_+}$, and $S^{x_-}$ \cite{Liu:2018gfc}. The critical points in this work are only related to the type $S^+$, as we consider every timelike tetrahedron to have at least one spacelike triangle \cite{Liu:2018gfc}. The expression of $S_{vef}=S^+_{vef}$ is given by\footnote{The expression is almost the same as the action derived using an alternative approach in \cite{Simao:2021qno}, with the only difference being the substitution of $l^+_{ef}$ with $l_{ef}^-$.}
\be
S_{vef}=j_f\left[2\ln\left(\sqrt{\frac{\langle l^{+}_{ef},Z_{vef}\rangle}{\langle Z_{vef}, l^+_{ef}\rangle}}\right)^{\kappa_{vef}}
-  \frac{\imath}{\gamma}\kappa_{vef}\ln\lt(\langle l^+_{ef},Z_{vef}\rangle \langle Z_{vef},l^+_{ef}\rangle\rt)\right].\label{timef}
\ee 
with the SU(1,1) invariant inner product $\langle\cdot,\cdot\rangle$. $l^-_{ef}$ does not appear in the action but is involved in the critical equation by $\mathring{z}_{vf}\propto \left(\mathring{\fg}^{\rm T}_{ve}\right)^{-1}\mathring{l}^{-}_{ef}$. 

In Figure \ref{fig:halfedgeaction}, we present the functions $\textbf{halfedgeaction}$ and $\textbf{halfedgeactiont}$ in the code. These functions compute the half-edge action for both the spacelike face and the timelike face using input data such as boundary data \textbf{xi}, the $\Slc$ variables \textbf{g}, the spinor variables \textbf{z}, orientation $\bm{\kappa}$, and the types of tetrahedrons and faces \textbf{meta}, \textbf{signzz}.
\begin{figure}
    \centering
    \includegraphics[scale=0.4]{figures/halfedgeaction.pdf}
    \caption{The computation of the half-edge action for the spacelike face and the timelike face}
    \label{fig:halfedgeaction}
\end{figure}

\subsection{Spinfoam action of 4-simplex and simplicial complex}

The action $S_v$ of the spinfoam 4-simplex amplitude is given by a sum of half-edge actions
\be
S_v=\sum_{(e,e')}\lt(S_{vef}+S_{ve'f}\rt),\label{SF}
\ee
where $(e,e')$ is an ordered pair of boundary tetrahdra (of the 4-simplex) sharing $f$. $S_{vef}$ is either \eqref{spacef} or \eqref{timef} depending on $f$ being spacelike or timelike. Both $e$ and $e'$ must be timelike when $f$ is timelike, whereas at least one tetrahedron between $e,e'$ is spacelike if $f$ is spacelike. For the 4-simplex amplitude, the integration variables are $\fg_{ve}$ and $z_{vf}$, while $j_f$, ${\xi}_{ef},\ \xi^\pm_{ef},\ l^+_{ef}$ are boundary data.

It is \textit{almost} straight-forward to construct the action $S$ on simplicial complex $\ck$ with more than one 4-simplex by 
\be
S=\sum_v S_v
\ee
where we sum over all 4-simplices $v$ in the complex. The only exception is for the internal spacelike triangle $h$ in a spacelike internal tetrahedron $e'$. The relevant terms in the above $S$ is $S_{ve'h}+S_{v'e'h}$ where $v,v'$ share $e'$. We make the following replacement \cite{Han:2013gna}
\be
S_{ve'f}+S_{v'e'f}\to j_h\lt[2  \ln \frac{\left\langle Z_{v e^{\prime} f}, Z_{v^{\prime} e^{\prime} f}\right\rangle}{\left\|Z_{v e^{\prime} f}\right\|\left\|Z_{v^{\prime} e^{\prime} f}\right\|}+\imath \gamma \ln \frac{\left\|Z_{v e^{\prime} f}\right\|^2}{\left\|Z_{v^{\prime} e^{\prime} f}\right\|^2}\rt] .\label{replacementspacef}
\ee
The orientation of $\partial f^*$ is outgoing from $v^*$ and incoming to $v'^*$. The SU(2) spinors $\xi_{ef}$ are excluded from the integration variables and can only be the boundary data, whereas the SU(1,1) spinors $\xi_{ef}^\pm,l_{ef}^+$ can still be the integration variables. The resulting action after the replacement \eqref{replacementspacef} is still denoted by $S$.

\subsection{Gauge freedom and gauge fixing in spinfoam action}

Since the spinfoam amplitude expressed in \eqref{SFamplitude} includes several types of continuous gauge degrees of freedom, it is necessary to introduce gauge fixings to eliminate the gauge degrees of freedom.
\bi
\item There is the $\Slc$ gauge transformation at each $v$:  
\be 
\fg_{ve}\mapsto x_v^{-1}\fg_{ve},\quad z_{vf}\mapsto x_v^{\rm T}z_{vf}, \quad x_v\in\Slc. 
\ee
To fix this gauge degree of freedom, we select one $\fg_{ve}$ to be a constant $\Slc$ matrix for each 4-simplex. The amplitude is unaffected by the choice of constant matrices.

\item For each $z_{vf}$, there is the scaling gauge freedom:
\be 
z_{vf}\mapsto\lambda_{vf} z_{vf}, \qquad \lambda_{vf}\in\mathbb{C}.  \label{zvf}
\ee 
We fix the gauge by setting one of the components of $z_{vf}$ to 1, i.e. either $z_{vf}=\left(1, {\a}_{vf}\right)^\mathrm{T}$ or $z_{vf}=\left({\a}_{vf},1\right)^\mathrm{T}$, where ${\a}_{vf}\in\C$. We adopt $z_{vf}=\left({\a}_{vf},1\right)^\mathrm{T}$, when the first component of $\mathring{z}_{vf}$ happens to be zero at the critical point.

\item At each spacelike triangle $h$ in an internal timelike tetrahedron, 
\be
\xi_{eh}^\pm\mapsto e^{i\psi_{eh}^\pm} \xi_{eh}^\pm,\qquad \psi^\pm_{eh}\in\mathbb{R},
\ee
leaves the action $S$ invariant. We fix the gauge by parametrizing 
\be
\xi^+_{eh}= \left(\cosh\theta_{eh},e^{-i\b_{eh}}\sinh\theta_{eh}\right)^{\rm T}, \quad \xi^-_{eh}= \left(e^{i\b_{eh}}\sinh\theta_{eh},\cosh\theta_{eh}\right)^{\rm T},\qquad \beta_{eh},\theta_{eh}\in\mathbb{R}.\label{xipm}
\ee 

\item At each internal timelike triangle $h$, the action $S$ is invariant under the scaling for $l^+_{eh}$
\be
l^+_{eh}\mapsto\lambda_{eh} l^+_{eh},\qquad \lambda_{eh}\in\mathbb{C}. \label{timelikegauge}
\ee
The scaling cancels between $F_{veh}$ and $F_{v'eh}$ for the orientation of $\partial h^*$ is outgoing from the vertex $v^*$ and incoming to $v'^*$, so it leaves the action $S$ invariant. Rigorously speaking, this scaling symmetry is not a gauge freedom if we parametrize $l^+_{eh}$ by $l^+_{eh}=v_{eh}\cdot (1,1)^{\rm T}$ with $v_{eh}\in \mathrm{SU}(1,1)$. But we still can use this symmetry to simplify the parametrization of $l^+_{eh}$. Namely, for any $l^+_{eh}=(l^{0+}_{eh},l^{1+}_{eh})^{\rm T}=v_{eh}\cdot (1,1)^{\rm T}$, we can modify $l^+_{eh}$ by $l^+_{eh}=(1,l^{1+}_{eh}/l^{0+}_{eh})^{\rm T} $ where $l^{1+}_{eh}/l^{0+}_{eh}\equiv e^{i\zeta_{eh}}$ is a phase implied by $v_{eh}\in \mathrm{SU}(1,1)$. Therefore we parametrize
\be
l^+_{eh}=(1,e^{i\zeta_{eh}})^{\rm T}
\ee
in the action, and $l_{eh}^-$ satisfies $\langle l^+_{eh},l^-_{eh}\rangle =1$ and $\langle l^-_{eh},l^-_{eh}\rangle =0$.

\item There is the $\Su$ gauge transformation for each bulk spacelike tetrahedron  $e$: 
\be 
\fg_{v'e}\mapsto \fg_{v'e}h_e^{\rm T},\quad \fg_{ve}\mapsto \fg_{ve}h_e^{\rm T}, \quad h_e\in\Su. \label{gaugesu2}
\ee 
To fix this $\Su$ gauge freedom, we parameterize either $\fg_{ve}$ or $\fg_{v'e}$ by a lower triangular matrix of the form
\begin{equation}
	k'=\left(\begin{matrix}
		\lambda^{-1}&0\\\mu &\lambda
	\end{matrix}\right),\ \lambda\in\mathbb{R}\setminus\{0\},\ \mu\in\mathbb{C} .\label{upper}
\end{equation} 
For any $g\in\Slc$, we can write $g=(g_0^\dagger)^{-1}$ and use the standard Iwasawa decomposition $g_0=ku$ where $k$ is an upper triangular matrix. Then we obtain $g=k'u$ where $k'=(k^\dagger)^{-1}$ is lower triangular. Choosing $h_e^{\rm T}=u^{-1}$ transforms $\fg_{v'e}$ to the lower triangular matrix $k'$.

\item In the examples considered in this paper, every timelike tetrahedron in our model contains at least one spacelike triangle and one timelike triangle (although timelike tetrahedron can have all faces spacelike). There is the $\mathrm{SU}(1,1)$ gauge transformation for each timelike tetrahedron $e$. 
\be 
\fg_{v'e}\mapsto \fg_{v'e}h_e^{\rm T},\quad \fg_{ve}\mapsto \fg_{ve}h_e^{\rm T}, \quad \xi_{eh}^\pm\mapsto h_e\xi_{eh}^\pm,\quad l_{eh}^\pm \mapsto  h_e l_{eh}^\pm,\quad 
\qquad h_e\in\mathrm{SU}(1,1). \label{gaugesu11}
\ee  
We implement the following procedure to fix this gauge freedom:  Firstly, We choose a spacelike triangle $f$ and find a $\mathrm{SU}(1,1)$ gauge transformation $U_e$ such that  	
    \be
	\xi^+_{ef}, \xi^-_{ef} \xRightarrow[]{U_{e}\in\mathrm{SU}(1,1)} \left(\begin{matrix}
		1\\
		0
	\end{matrix}\right), \,\left(\begin{matrix}
		0\\
		1
	\end{matrix}\right). 
	\ee 
The corresponding $U_e$ acts on all $\xi^{\pm}_{ef}$ or $l^\pm_{eh}$ in this tetrahedron. Secondly, choose a timelike triangle $h$ and rewrite the corresponding $l^+_{eh}=(e^{-i\zeta_{eh}/2},e^{i\zeta_{eh}/2})^{\rm T}$ by the scaling symmetry.  
 
Then we make a futher $\mathrm{SU}(1,1)$ gauge transformation 
	\be
	\tilde{U}_e= \left(\begin{matrix}
		e^{i\zeta_{eh}/2} & 0\\
		0& e^{-i\zeta_{eh}/2}
	\end{matrix}\right). 
	\ee 
This matrix $\tilde{U}_e$ fixes $l^+_{eh}$ to $(1,1)^{\rm T}$. 
 
This $\tilde{U}_e$ again acts on all $\xi_{ef}^\pm$ and $l_{eh}^\pm$ within the same tetrahedron. In particular, for the spacelike $f$, $\xi_{ef}^\pm$ become
 $	 \xi^+_{ef}= (	e^{i{\zeta_{eh}}/{2}},0)^{\rm T}, \   \xi^-_{ef} = (0, e^{-i{\zeta_{eh}}/{2}})^{\rm T}$, whereas the phases can be further removed by the gauge transformation of $\xi_{ef}^\pm$. These gauge transformations allow us to gauge fixing the timelike face $l^\pm_{eh}$ in the form: 
\be
l^+_{eh}= \left(\begin{matrix}
		1\\1
	\end{matrix}\right), \qquad 
\xi^+_{ef}= \left(\begin{matrix}
		1\\
		0
	\end{matrix}\right), \qquad
\xi^-_{ef}=\left(\begin{matrix}
		0\\
		1
	\end{matrix}\right)\label{GFform}
	\ee 
\item The spinfoam action also has the following discrete gauge symmetry: flipping the sign of the group variable, $\frak{g}_{ve} \rightarrow - \frak{g}_{ve}$. Therefore, the space of group variables essentially corresponds to the restricted Lorentz group $\mathrm{SO}^+(1,3)$ rather than its double-cover $\Slc$.

\end{itemize}

For a generic data $(\fg_{ve},z_{vf},\xi_{eh}^{\pm},l^+_{eh})$, firstly we can use the SU(1,1) gauge transform to fix $\xi_{ef}^{\pm}$ and $l_{eh}^+$ into the gauge fixing form \eqref{GFform} for all timelike internal tetrahedra. Secondly, we use the $\mathrm{SL}(2,\C)$ gauge transformation fix one $\fg_{ve} $ to be constant within each 4-simplex. In our model, every 4-simplex has at least one timelike tetrahedron. We always choose the gauge fixed $\fg_{ve}$ to associate with the timelike $e$, if $v$ does not have boundary tetrahedron, or with the boundary $e$, if $v$ has the boundary tetrahedron. Since the $\mathrm{SL}(2,\C)$ gauge transformation acts to the left of $\fg_{ve}$, it does not affect the SU(1,1) gauge fixing \eqref{GFform}. Thirdly, we use the SU(2) gauge transformation to put one of $\fg_{ve}$ to lower triangular matrix for each spacelike internal $e$. The SU(2) gauge transformation does not affect the previous gauge fixings, since it only acts on spacelike tetrahedra and acts to the right of $\fg_{ve}$. Lastly, we use the scaling symmetry to reduce $z_{vf}, \xi^{\pm}_{eh},l^+_{eh}$ to the gauge fixing forms. This procedure can transform any data $(\fg_{ve},z_{vf},\xi_{eh}^{\pm},l^+_{eh})$ to the gauge fixing form defined above, so it shows the above gauge fixing is well-defined.

\subsection{Poisson Summation}
We would like to change the sum over $j_h$ in Eq. (\ref{SFamplitude}) to the integral, preparing for the stationary phase analysis. The idea is to apply a generalization of the Poisson summation formula \cite{BJBCrowley_1979}
\[
\sum_{n=0}^{n_{\rm max}} f(n)=\sum_{k \in \mathbb{Z}} \int_{-\epsilon}^{n_{\rm max}+1-\epsilon} \mathrm{d} n f(n) \,{e}^{2\pi i k n},
\]
where $\epsilon\in(0,1)$. Identifying $n=2j$, $n_{\rm max}=2j^{\rm max}$ and $f(n)$ to be the summand in the spinfoam amplitude, we obtain the following expression of $A(\ck)$
\be
A(\ck)&=&\sum_{\{k_h\in\mathbb{Z}\}}\int\limits_{-\epsilon}^{2j^{\rm max}+1-\epsilon} 
\prod_f\mathrm{d}(2 j_{f}) \,\prod_h\mu_h(j_f)\int [\rmd X]\, e^{ S_{\rm SF}^{(k)}}, \label{integralFormAmp1}\\
&&S_{\rm SF}^{(k)}=S_{\rm SF}+4\pi i \sum_f j_f k_f
\ee
To probe the large-$j$ regime, we scale boundary spins $j_b\to \l j_b$ with any $\l\gg1$, and make the change of variables $j_h\to\l j_h$. We also scale $j^{\rm max}$ by $j^{\rm max}\to \l j^{\rm max}$. Then, $A(\ck)$ is given by  
\be
A(\ck)&=&\sum_{\{k_h\in\mathbb{Z}\}}\int\limits_{-\epsilon/\lambda }^{2j^{\rm max}+({1-\epsilon})/{\l}} 
\prod_f\mathrm{d}(2 j_{f}) \,\prod_h\mu_h(\l j_f)\int [\rmd X]\, e^{\l S_{\rm SF}^{(k)}}\label{integralFormAmp22}
\ee
which is used in our discussion. Here, we focus on the amplitude with all $k_h=0$, as this configuration often corresponds to the dominant contribution, especially under suitable boundary conditions, as shown in the numerical results in e.g., \cite{Han:2021kll}.

\section{Boundary data and critical point of a 4-simplex amplitude}\label{v1data}

\begin{figure}[h]
	\centering
	\includegraphics[scale=0.12]{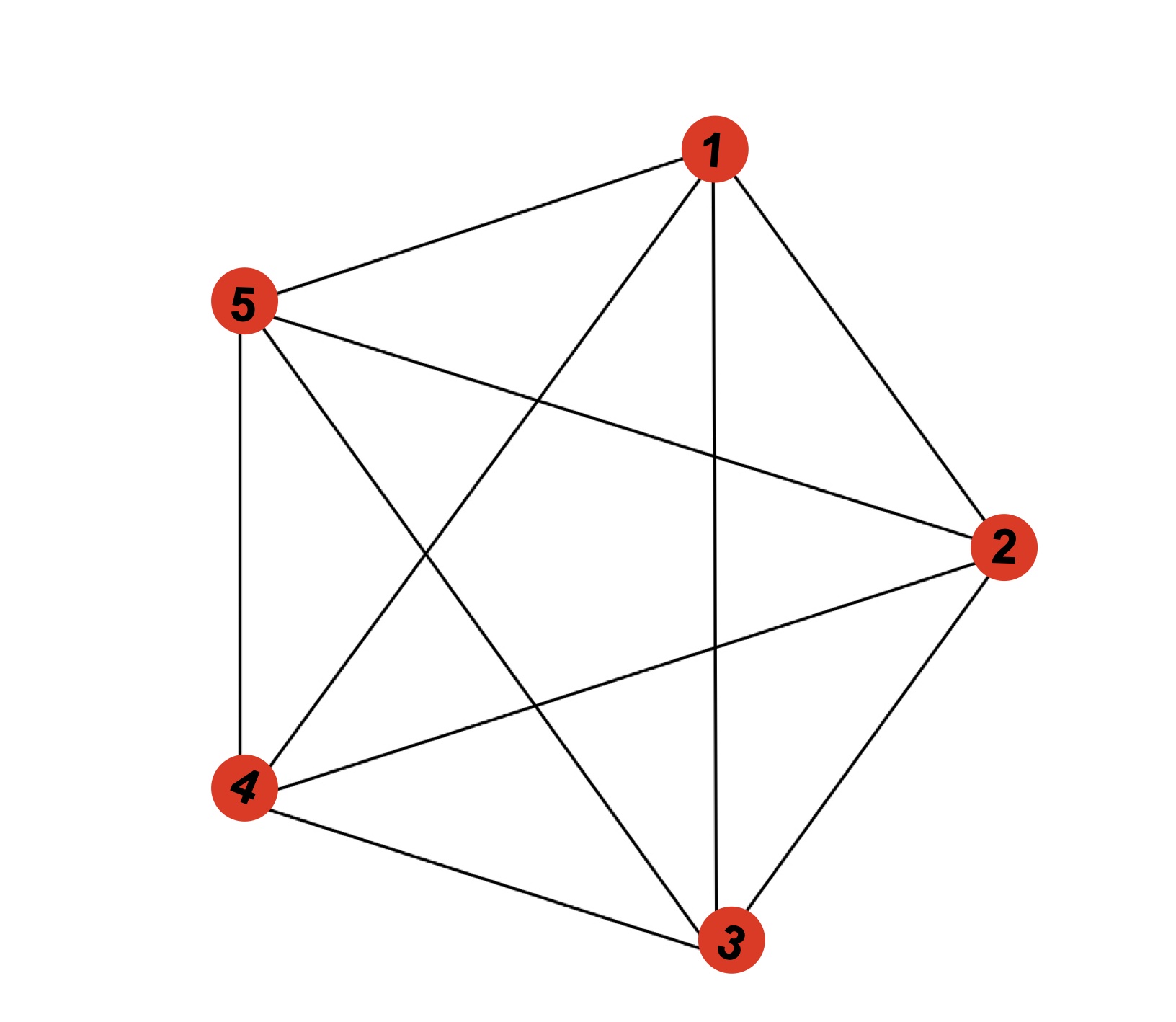}\label{4simplex}
\end{figure}

A 4-simplex $v$ has five vertices and ten edges connecting every vertices to all other vertices. Given a 4-simplex with vertices $(1,2,3,4,5)$ (see Figure \ref{4simplex}), all sub-simplices can be labelled by a subset of vertices, e.g. $(1,2,3,4)$ labels a tetrahedron, $(1,2,3)$ labels a triangle, and $(1,2)$ labels an edge, etc.

\subsection{Four-dimensional outgoing normals}

\begin{figure}[ht]
    \centering
    \caption{ Computation of 4d normals}
    \includegraphics[scale=0.5]{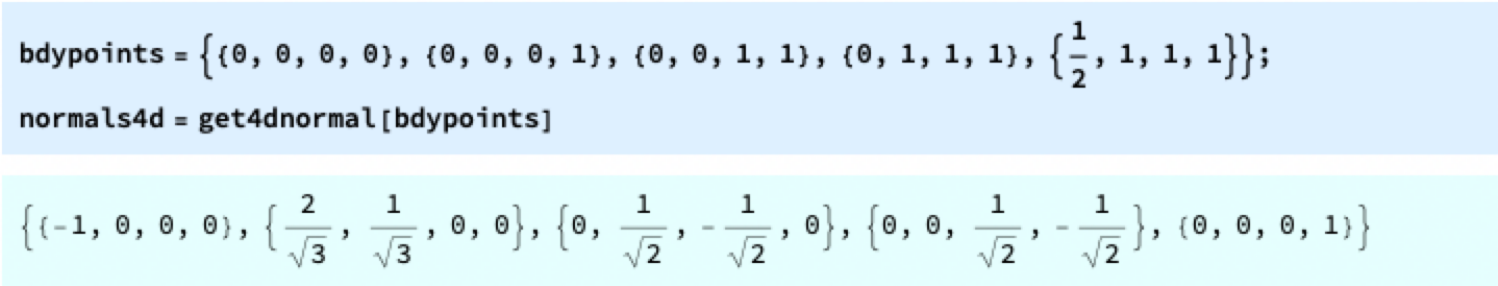}
    \label{fig:4dnormals}
\end{figure}

To specify a flat 4-simplex geometry numerically, we use the coordinates $(t,x,y,z)$ of its vertices in $(\R^4,\eta)$. As an example, the 4-simplex geometry is determined by the following coordinates for vertices $P_1,\cdots,P_5$:
\be
P_1 = (0,0,0,0),\quad
P_2 = (0,0,0,1),\quad
P_3 = (0,0,1,1),\quad
P_4 = (0,1,1,1),\quad
P_5 = (\frac{1}{2},1,1,1).\label{vertices}
\ee 
These coordinates correspond to the vertices of a 4-simplex with squared edge lengths  
\be
s^2_{ij} =\left(1,2,3,\frac{11}{4},1,2,\frac{7}{4},1,\frac{3}{4},-\frac{1}{4}\right), \label{lengthsq}
\ee 
where $s_{ij}^2 > 0$ corresponds to a spacelike edge, while $s_{ij}^2 < 0$ corresponds to a timelike edge. With the given coordinates, we can compute the edge vectors $P_{ij} = P_i - P_j$ for each pair of vertices. For each of the five tetrahedra $e$ \footnote{The tetrahedra are $\{e_1,e_2,e_3,e_4,e_5\}=\{(1,2,3,4),\, (1,2,3,5),\, (1,2,4,5),\, (1,3,4,5),\, (2,3,4,5)\}$.}, the four-dimensional normal vector $N_{ve}$ is perpendicular to all edges of $e$, meaning that 
\be
N_{ve}\cdot P_{ij} = 0,\qquad\forall\ P_{ij}\subset e.
\ee
Here, the dot product is taken with respect to the metric $\eta=(-1,1,1,1)$. There are only 3 independent $P_{ij}$, which corresponds to 3 equations determining $N_{ve}$ up to scaling. Then we normalize each vector to obtain a unit vector. However, the resulting unit vector can be either outgoing or incoming. As we know, the 4-simplex with the coordinates $P_i$ in (\ref{vertices}) spans a convex hull:
\be 
\sigma = \left\{\sum_{i=1}^5 t_k P_i; \quad t_i\geq 0, \quad \sum_{i=1}^5 t_i=1 \right\}.
\ee 
If $N_{ve}$ is an outgoing normal originating from the barycenter $\fp_e$ of the tetrahedron $e$, $\fp_e+\epsilon N_{ve}$ for any $\epsilon>0$ is a point lying outside $\sigma$. With these constraints, we can subsequently select the corresponding outgoing 4-dimensional normals. This computation is implemented by the function \textbf{get4dnormal} in Mathematica. In FIG.\ref{fig:4dnormals}, we illustrate the computation of normalized four-dimensional normals $N_{ve}$ using the input of \textbf{bdypoints} and the corresponding normalized 4d normals for this particular 4-simplex.

Some normals in FIG. \ref{fig:4dnormals} are spacelike, while others are timelike. The spacelike tetrahedra have 4d timelike normals, and all of their faces are spacelike. The timelike tetrahedra have 4d spacelike normals, and their faces can be either spacelike or timelike. {In this example, each timelike tetrahedron consistently consists of two spacelike faces and two timelike faces.} We introduce $\sgn(e)$ to label spacelike tetrahedra as ``+1'' and timelike tetrahedra as ``-1'' (as shown in Table \ref{sgndet}).

\begin{table}[H]
	\centering\caption{The types of each tetrahedron \textbf{sgndet} are indicated by either ``+1'' for spacelike tetrahedra or ``-1'' for timelike tetrahedra. The vertex coordinates for each tetrahedron are listed in \eqref{vertices}.\\ }\label{sgndet}
	\begin{tabular}{|c|c|c|c|c|c|}
		\hline
		$e$&$e_1$&$e_2$&$e_3$&$e_4$&$e_5$\\
		\hline
		 \textbf{sgndet} &+1&+1&-1&-1&-1\\
		\hline
	\end{tabular}
\end{table}

The closure condition (of the 4-simplex) satisfied by the outgoing normal $N_{ve}$ is 
\be
\sum_{i=1}^5W_iN_{ve_i}=0
\ee
where $W_i$ is proportional to the volume of $e_i$ and $W_i>0$ $(W_i<0)$ for spacelike (timelike) tetrahedron.

\subsection{Dihedral angles} \label{dihedralAnglesSec}
\begin{figure}[h]
    \centering
    \caption{Computation of dihedral angles in a 4-simplex}
    \includegraphics[scale=0.5]{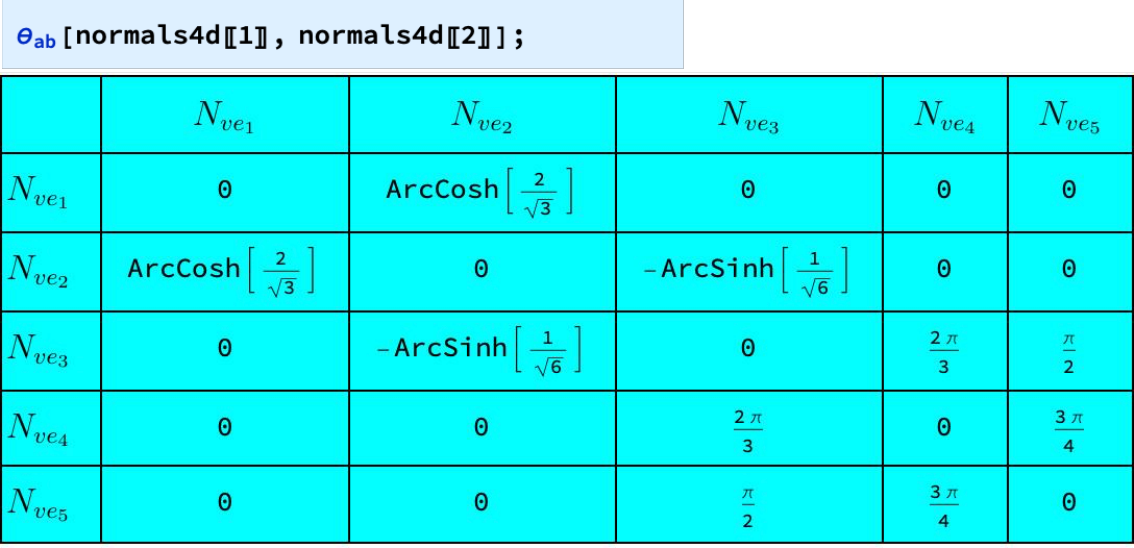}
    \label{fig:dihedrals}
\end{figure}

In the 4-simplex (\ref{vertices}), it includes spacelike and timelike normals, we adopt the following definitions to determine dihedral angles of two {outgoing} normals $N_{ve},N_{ve'}$: 
\begin{itemize}
	\item Lorentzian signature $(-+++)$ and both $N_{ve}$ and $N_{ve'}$ are timelike, the dihedral angle is:
	\be
	\theta_{e,e'} = \begin{cases}\cosh^{-1} \left(N_{ve}\cdot N_{ve'}\right), & \text { if } N_{ve} \cdot N_{ve'}>0 \\ 
    -\cosh^{-1}\left(-N_{ve} \cdot  N_{ve'}\right), & \text { if } N_{ve}\cdot N_{ve'}<0\end{cases}
	\ee  
	\item In cases where one of $N_{ve}$ and $N_{ve'}$ is spacelike and the other is timelike with Lorentzian signature $(-+++)$, the dihedral angle is determined as:
	\be
	\theta_{e,e'}=-\sinh^{-1}\left(N_{ve} \cdot N_{ve'}\right).
	\ee 
	\item When both $N_{ve}$ and $N_{ve'}$ are spacelike with Euclidean signature $(++++)$, the dihedral angle is computed as follows:
	\be
 \theta_{e,e'}=\cos^{-1}\left(N_{ve}\cdot N_{ve'}\right)>0.
    \ee
\end{itemize}
In FIG. \ref{fig:dihedrals}, we show the function $\bm{\theta}_{\bf ab}$ for computing the dihedral angles with the input \textbf{normals4d} and the numerical results from the data in FIG. \ref{fig:4dnormals}.  

\subsection{SO(1,3) and SL(2,$\mathbb{C}$) group elements}

The real critical points of the spinfoam model can be constructed from the geometrical data. Let us firstly set up some notations. A bivector $B^{IJ}$ in Minkowski spacetime is an antisymmetric tensor with two upper indices $I,J=,0,\cdots,3$. A pair of 4d vectors $N,\,M$ determine a simple bivector $N\wedge M$ by
\be 
B=N\wedge M = N\otimes M - M\otimes N. 
\ee 
The norm $\left| B\right|$ of a bivector $B$ is defined as
\be 
 \left| B\right|^2 = \frac{1}{2} \lt|B^{IJ}B_{IJ}\rt|\ .
\ee 
The indices are raised and lowered with the Minkowski metric $\eta$. 

We choose a reference 4d normal for each tetrahedron $e$ by $N_{\rm{ref}}=(\pm 1,0,0,0)^T$, if $e$ is spacelike with $N_{ve}$ future/past-pointing, or by $N_{\rm{ref}} = (0,0,0, 1)^T$, if $e$ is timelike. We would like to use the geometrical data of the 4-simplex to determine $\Lambda_{ve}\in \rm{SO(1,3)}$ satisfying
\be 
\Lambda_{ve}\, N_{\rm{ref}}  = N_{ve} .\label{LambdaNN}
\ee  
We construct $\Lambda_{ve}$ by using the outgoing normal $N_{ve}$ \cite{Barrett:2009gg,Kaminski:2017eew,Han:2011re}\footnote{In the case of timelike $N_{ve}$, without losing generality, we choose the plane spanned by $N_{ve}$ and $N_{\rm ref}$ to be the $t$-$x$ plane, then we can parametrize $N_{ve}=\pm(\cosh (\theta ),-\sinh (\theta ),0,0)$. We verify that $\mathbf{B}_{ve}=-\sgn(\theta)K^1$ and $N_{ve}=\exp(|\theta| \mathbf{B}_{ve})N_{\rm ref}$ where $N_{\rm ref}=\pm(1,0,0,0)$. Since $\cosh(\theta)=-N_{\rm ref}\cdot N_{ve}=\cosh(\theta_{\mathrm{ref},e})$, we obtain $|\theta| =|\theta_{\mathrm{ref},e}|=-\theta_{\mathrm{ref},e}$.

In the case of spacelike $N_{ve}$, we choose the plane spanned by $N_{ve}$ and $N_{\rm ref}$ to be the $y$-$z$ plane, then we can parametrize $N_{ve}=(0,0,\sin (\theta ),\cos(\theta ))$. We verify that $\mathbf{B}_{ve}=-\sgn(\theta)J^1$ and $N_{ve}=\exp(-|\theta| \mathbf{B}_{ve})N_{\rm ref}$ where $N_{\rm ref}=\pm(0,0,0,1)$. Since $\cos(\theta)=N_{\rm ref}\cdot N_{ve}=\cos(\theta_{\mathrm{ref},e})$, we obtain $|\theta| =|\theta_{\mathrm{ref},e}|=\theta_{\mathrm{ref},e}$.},
\be
\Lambda_{ve} =\exp\left(\Theta_{\text{ref}, e}\,\mathbf{B}_{ve}\right),\qquad \mathbf{B}_{ve}= \frac{N_{\rm{ref}}\wedge N_{ve}}{\left| N_{\rm{ref}}\wedge N_{ve}\right|},\qquad \Theta_{\text{ref}, e}=-\theta_{\text{ref}, e} \label{so13property}
\ee 
where $\theta_{\text{ref}, e}$ is the dihedral angle between the reference normal and the 4d normal defined in (\ref{dihedralAnglesSec}). Since we always have $N_{\rm ref}\cdot N_{ve}<0$ for timelike $N_{ve}$ and $N_{\rm ref}\cdot N_{ve}>0$ for spacelike $N_{ve}$
\be
\begin{cases} \theta_{\text{ref},e}<0, & \text { if } {N_{ve}} \text { is timelike normal,}\\ \theta_{\text{ref}, e}>0, & \text { if } {N_{ve}} \text{ is spacelike normal.}\end{cases}
\ee

\begin{figure}[ht]
    \centering
    \caption{Computation of $\rm SO (1,3)$ solution}
    \includegraphics[scale=0.45]{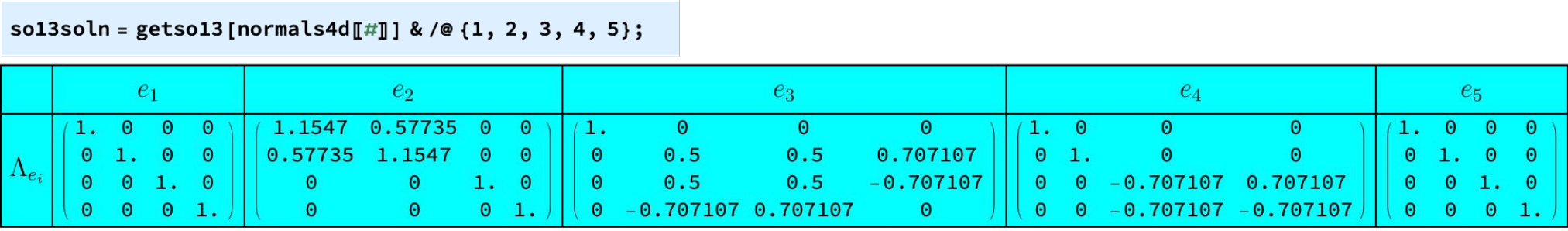}
    \label{fig:so13soln}
\end{figure}

The function of computing $\L_{ve}$ is \textbf{getso13}. In FIG. \ref{fig:so13soln}, we show how to compute $\rm SO (1,3)$ solutions with the input 4d normals \textbf{normals4d} and the numerical results. 

\begin{figure}[h]
    \centering
    \caption{Computation of $\Slc$ solution}
    \includegraphics[scale=0.5]{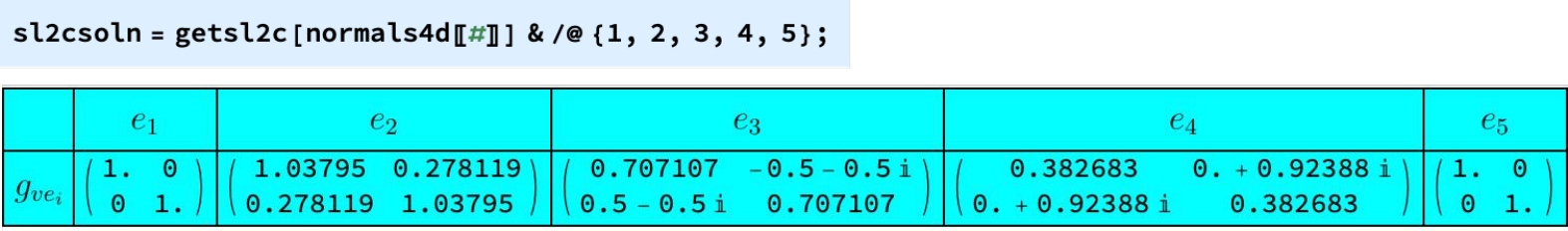}
    \label{fig:sl2csoln}
\end{figure}

The normalized bivector $\mathbf{B}_{ve}$ in (\ref{so13property}) can be identified as an element in the Lorentz Lie algebra $so(1,3)$. We choose the basis of $so(1,3)$ by:
\be
{\begin{aligned}K^{1}&={\begin{pmatrix}0&1&0&0\\1&0&0&0\\0&0&0&0\\0&0&0&0\end{pmatrix}}, & K^{2}&={\begin{pmatrix}0&0&1&0\\0&0&0&0\\1&0&0&0\\0&0&0&0\end{pmatrix}},& K^{3}&=\begin{pmatrix}0&0&0&1\\0&0&0&0\\0&0&0&0\\1&0&0&0\end{pmatrix},\\
J^{1}&={\begin{pmatrix}0&0&0&0\\0&0&0&0\\0&0&0&1\\0&0&-1&0\end{pmatrix}}, & J^{2}&={\begin{pmatrix}0&0&0&0\\0&0&0&-1\\0&0&0&0\\0&1&0&0\end{pmatrix}}, &J^{3}&={\begin{pmatrix}0&0&0&0\\0&0&1&0\\0&-1&0&0\\0&0&0&0\end{pmatrix}}.\\[8pt]\end{aligned}}\label{JandK}
\ee  
The commutation relations are
\be 
\left[K^{i},K^{j}\right]=\epsilon^{ijk}J^{k},\quad \left[J^{i},J^{j}\right]=-\epsilon^{ijk}J^{k},\quad\left[J^{i},K^{j}\right]=-
\epsilon^{ijk}K^{k}. 
\ee 
The $4\times 4$ matrices in \eqref{JandK} are the spin-1 representations of the Lie algebra generators $J^i,K^i$. The spin-$1/2$ representations of $J^i,K^i$ are given by 
\be 
\pi(K^i) = \frac{\sigma^i}{2},\quad \pi(J^i) = \frac{\imath \sigma^i}{2}, \quad \text{where}\quad i=1,2,3,
\ee 
Any bivector $\mathbf{B}_{ve}$ is a linear combination of $(K^i,\, J^i)$ basis: 
\be 
\mathbf{B}_{ve}=\sum_{i=1}^{3}\alpha_i K^i+\sum_{j=1}^{3}\beta_j J^j, \quad \alpha_i = \frac{1}{2}\Tr(\mathbf{B}_{ve} K_i),\quad \beta_i = -\frac{1}{2}\Tr(\mathbf{B}_{ve}J_i). \label{bfXa}
\ee 
The $4\times 4$ bivector $\mathbf{B}_{ve}$ has the spin-$1/2$ representation, denoted by $B_{ve}$, 
\be 
B_{ve} = \sum_{i=1}^3\alpha_i \pi(K^i) + \sum_{i=3}^3\beta_i \pi(J^i). \label{2dBivector}
\ee 
The previously deﬁned $\Lambda_{ve}\in\rm{SO}(1,3)$ lifts to $g_{ve}\in \Slc$ given  by
\be 
g_{ve} = \pm\exp\left(\Theta_{\text{ref}, e}\,B_{ve}\right).
\ee
The sign $\pm$ for each $g_{ve}$ are the discrete gauge freedom of \eqref{SFamplitude}. We fix the gauge by choosing the $+$ sign. The resulting $g_{ve}$ is one of the spinfoam data at the critical point. The function of computing $g_{ve}$ is \textbf{getsl2c}. In FIG. \ref{fig:sl2csoln}, we show how to compute $\Slc$ solutions with the input 4d normals and the numerical results. The $\Slc$ solutions $g_{ve}$ computed here and the $\Slc$ group element $\fg_{ve}$ in the spinfoam action (\ref{SF}) have the relation\footnote{This convention is the same as in \cite{Kaminski:2017eew} for spacelike triangles.}:
\be
\mathring{\fg}_{ve}=(g_{ve}^{\rm T})^{-1}, \label{gvegve}
\ee
where $\mathring{\fg}_{ve}$ denotes $\fg_{ve}$ evaluated at the real critical point. The data of $\mathring{\fg}_{ve}$ is stored in \textbf{gdataof} in the code.

One can check that $g_{ve}\in \Slc$ satisfies
\be
g_{ve} \,H_{\rm ref}\, g_{ve}^{\dagger} = H_{ve}, \label{gHgH}
\ee 
which translates \eqref{LambdaNN} to the spin-$1/2$ representation. $H_e$ is the Hermitian matrix that is converted from 4d normals $N_{ve} = (t,x,y,z)$ by the following
\be
H_{ve} =t\mathbb{1} + x\sigma_1 + y \sigma_2 + z \sigma_3= \left(\begin{matrix}
	t+z & x- \imath y\\
	x+ \imath y & t-z
\end{matrix}\right).
\ee 
The reference Hermitian matrix $H_{\text{ref}}$ can take on two possible values depending on the choice of reference normal. If the tetrahedron is spacelike, and the reference normal is $(\pm1,0,0,0)$, then $H_{\text{ref}}$ is $\left(\begin{matrix}
	\pm1&0\\
	0&\pm1
\end{matrix}\right)$. If the tetrahedron is timelike, and the reference normal is $(0,0,0,1)$, then $H_{\text{ref}}$ is $\left(\begin{matrix}
	1&0\\
	0&-1
\end{matrix}\right)$.

We define $\bar{\sigma}_I=-(\mathbb{1},\sigma_i)$. By raising the index, $\bar{\sigma}^I=(\mathbb{1},-\sigma^i)$ comparing with $\sigma^I=(\mathbb{1},\sigma^i)$ (we have $\sigma^i=\sigma_i$ and $\sigma^I=-\bar{\sigma}_I$). any 4d vector $V$ relates to the Hermitian matrix $H_{V}$ by $H_{V}=-V^I\bar{\sigma}_I$. Eq.\eqref{gHgH} is implied by 
\be
g \bar{\sigma}_J g^\dagger =\bar{\sigma}_I\L^I_{\ J}, 
\ee
which relates $g\in\Slc$ to $\L\in \mathrm{SO}^+(1,3)$. Then the SO(1,3) matrix $\L$ is given by 
\be
\L^I_{\ J}=\frac{1}{2}\Tr(\bar{\sigma}_I g \bar{\sigma}_J g^\dagger)=\frac{1}{2}\Tr({\sigma}^I g {\sigma}^J g^\dagger),\quad \text{or equivalently}\quad (\L^{-1})^I_{\ J}=\frac{1}{2}\Tr(g\bar{\sigma}^I g^\dagger \bar{\sigma}^J ).\label{Lambdagg}
\ee

\subsection{Triangle areas $A_f$}
\begin{figure}[h]
    \centering
    \includegraphics[scale=0.3]{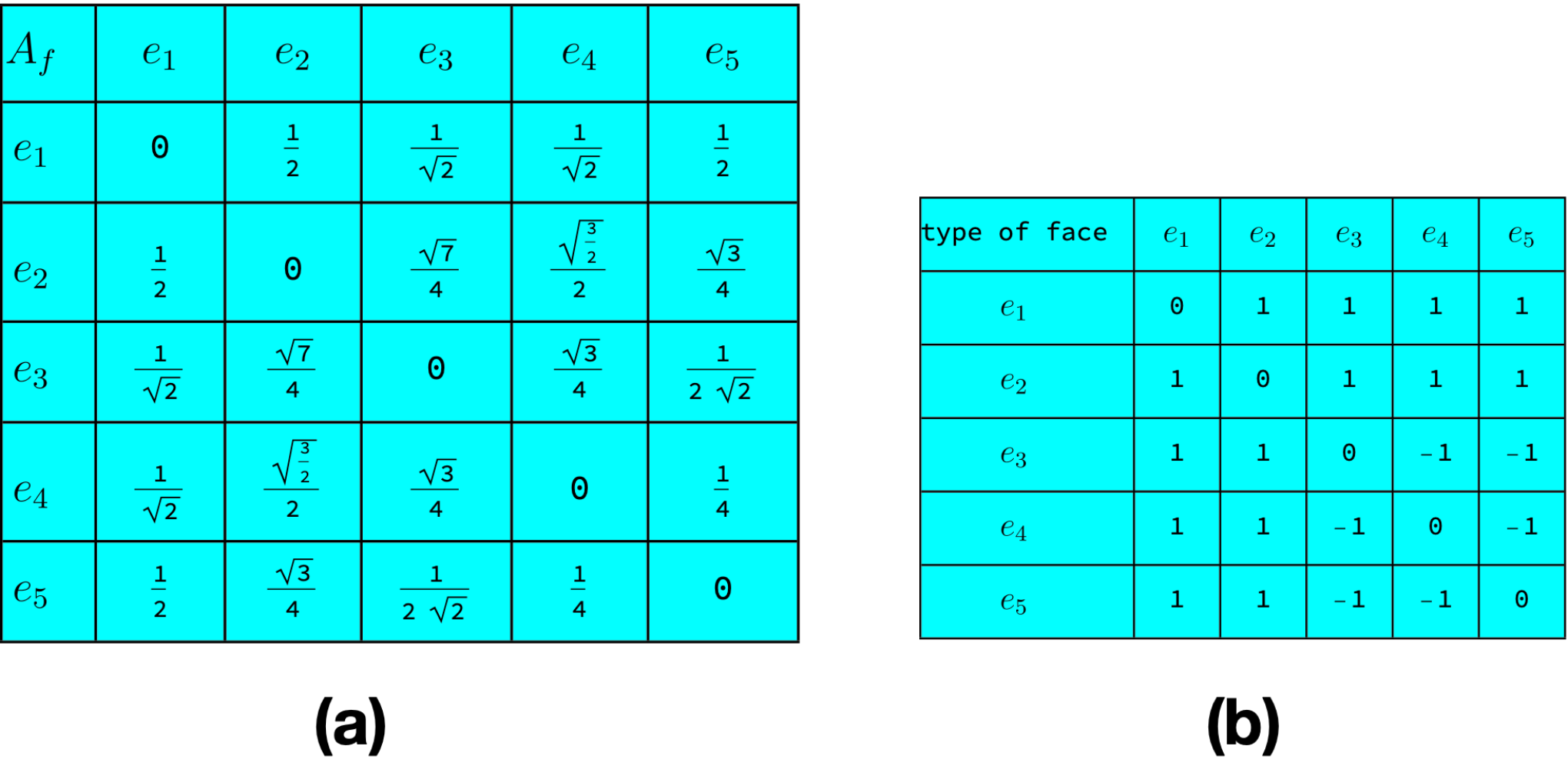}
    \caption{Each cell of the table is the area and the type for the face shared by the tetrahedron in row and the tetrahedron in column.}
    \label{fig:areas}
\end{figure}

The signed squared volume of a $d$-dimensional simplex $\Fs^d = (0\cdots d)$ with vertices $\{0,\dots,d\}$ can be computed via the determinant of the associated Caley-Menger matrix for the signed squared edge lengths~\cite{Tate:2011rm},
\be\label{eq::VolumeCaleyMenger}
\mathbb{V}_{\Fs^d}&\,=\,& -\frac{ (-1)^{d} }{ 2^{d} (d!)^2} \det  \left(
\begin{matrix}
	0 & 1 & 1 & 1 & \cdots & 1 \\
	1 & 0 & s_{01} & s_{02} & \cdots & s_{0d} \\
	1 & s_{01} & 0 & s_{12} & \cdots & s_{1d} \\
	\vdots & \vdots & \vdots & \vdots & \ddots & \vdots \\
	1 & s_{0d} & s_{1d} & s_{2d} & \cdots & 0
\end{matrix}
\right).
\ee
We compute the signed squared area $\mathbb{A}_f$ of the triangle $f$, with $d=2$. Note that $\mathbb{A}_f<0$ means a timelike triangle and  $\mathbb{A}_f>0$ means a spacelike triangle. The area is given by
\be 
A_f = \sqrt{\left|\mathbb{A}_f\right|}=\begin{cases}
    \gamma j_f, & \text{spacelike triangle}\\
    j_f, & \text{timelike triangle}
\end{cases}.
\ee 
where $j_f\in \mathbb{N}_0/2$. In FIG. \ref{fig:areas} (a), we show the numerical results of areas. FIG. \ref{fig:areas} (b) shows the types of each triangle shared by tetrahedra $e_i$ and $e_j$: $+1$ labels a spacelike face, and $-1$ labels a timelike face. Although the areas in FIG. \ref{fig:areas} correspond to non-half-integer $j_f$, by multiplying $j_f$ by $\l\gg 1$, we can make $\l \lt[j_f+O(1/\l)\rt]$ a half-integer, where $j_f+O(1/\l)$ truncates the decimal expansion of $j_f$ to certain order \footnote{For example, $10^6\lt(\frac{1}{\sqrt{2}}-\frac{6.78119\cdots}{10^{6}}\rt)=707100$.}. $O(1/\l)$ is subleading in the stationary phase approximation. 

\subsection{Spinors $\xi_{ef}$ and $l_{ef}^\pm $}\label{xicomputation}

\begin{figure}[ht]
    \centering
    \caption{Computation and results of $P^{(4)}_{e,I}$.}
    \includegraphics[scale=0.5]{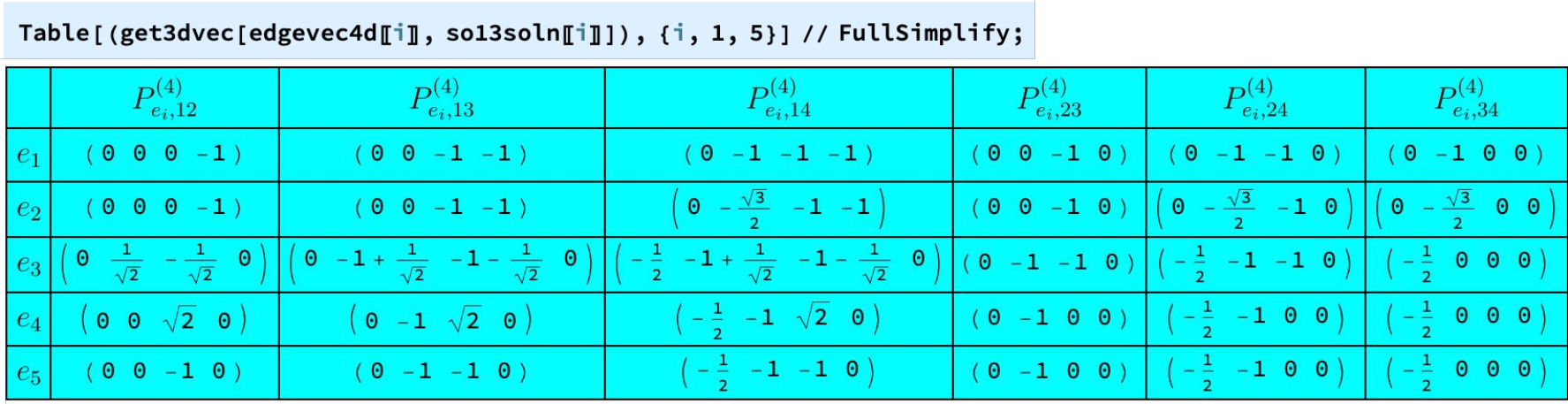}
    \label{fig:4d edgevec}
\end{figure}

For each tetrahedron $e$ and in its own reference frame where its 4d normal is $N_{\rm ref}$, their edge vectors are obtained by
\be 
P_{e,ij}^{(4)} = \Lambda^{-1}_{ve} P_{ij}, \quad i,j=1,\cdots,5. 
\ee 
Here, $P_{ij}=P_i-P_j$ is the 4d edge vector expressed in terms of (\ref{vertices}). If $e$ is spacelike (timelike),  the first (last) component of $P_{e, ij}^{(4)}$ is 0. This process is repeated for all edge vectors in all tetrahedra to obtain the 3d edge vectors that describe the geometry of the entire triangulation. In FIG. \ref{fig:4d edgevec},  we demonstrate the use of the function \textbf{get3dvec} to compute $P^{(4)}_{e,ij}$ with the input of 4D edge vectors \textbf{edgevec4d} and the SO(1,3) solutions \textbf{so13soln} for each tetrahedron, along with presenting the numerical results.

\begin{figure}[h]
    \centering
    \caption{Computation of $X_{ef}$}
    \includegraphics[scale=0.45]{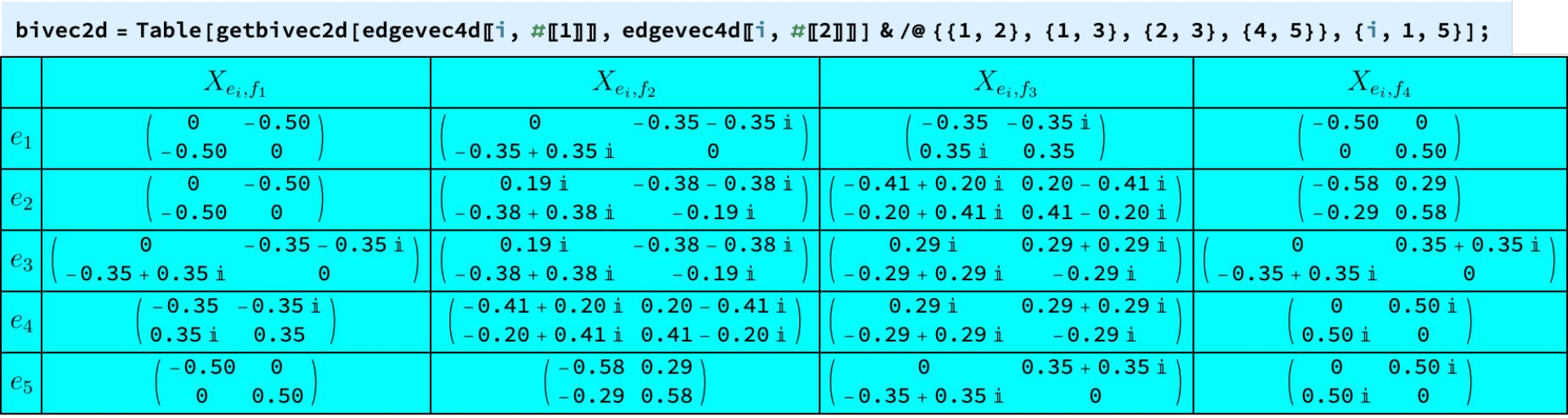}
    \label{fig:Bivec}
\end{figure}

The bivector $\mathbf{X}_{ef}$ of the triangle $f$ in the tetrahedron $e$ with vertices $(P_i,\,P_j,\,P_k)$ is given by 
\be
\mathbb{X}^{IJ}_{e f}=\frac{1}{2}\epsilon^{IJ}_{\ \ KL}\, P_{e, ij}^{(4)\ K} P_{e, ik}^{(4)\ L}.
\ee
Then, the normalized bivector is
\be 
\mathbf{X}_{ef} =\frac{\mathbb{X}_{ef}}{|\mathbb{X}_{ef}|}
\ee
We can convert $\mathbf{X}_{ef}$ from the spin-1 representation to the spin-$1/2$ representation by 
\be 
X_{ef} = \sum_{i=1}^3\alpha_i \pi(K^i) + \beta_i\pi(J^i), \quad \text{where}\quad \alpha_i= \frac{1}{2}\mathtt{Tr}(\mathbf{X}_{ef}K^i),\quad \beta_i= -\frac{1}{2}\mathtt{Tr}(\mathbf{X}_{ef}J^i). 
\ee
In FIG. \ref{fig:Bivec},  we demonstrate the use of the function \textbf{getbivec2d} for computing $X_{ef}$ with the input of 2 edge vectors $P^{(4)}_{e,ij}$ and $P^{(4)}_{e,kl}$ and demonstrate the corresponding numerical results.

It can be verified that the bivectors in the spin-1/2 representation $X_{ef}$ obtained in FIG. \ref{fig:Bivec} satisfy the parallel transport, expressed as:
\be 
g_{ve}X_{ef} g^{-1}_{ve} = g_{ve'} X_{e'f} g^{-1}_{ve'}
\ee 
which is equivalent to 
\be 
\left(\mathring{\fg}^{\rm T}_{ve}\right)^{-1} X_{ef} \mathring{\fg}^{\rm T}_{ve} = \left(\mathring{\fg}^{\rm T}_{ve'}\right)^{-1} X_{e'f} \mathring{\fg}^{\rm T}_{ve'}.
\ee 
This indicates that the data that we construct relate to the spinfoam critical point, since the critical equations of spinfoam amplitude result in the parallel transport of bivectors \cite{Han:2011re,Kaminski:2017eew,Liu:2018gfc}.

We can now use the bivector $X_{ef}$ to construct the corresponding $\Su$ or $\text{SU}(1,1)$ spinors. The spacelike triangle associates with the SU(2) or SU(1,1) spinor $\xi_{ef}$, depending on the tetrahedron is spacelike or timelike. The timelike triangle associates with the SU(1,1) spinor $l_{ef}^\pm $. The spinor is given by an SU(2) or SU(1,1) matrix $S_{\rm SU(2)}$ or $S_{\rm SU(1,1)}$ acting on certain reference spinor. In the following, we discuss all three cases: spacelike triangle in spacelike tetrahedron,  spacelike triangle in timelike tetrahedron, timelike triangle in timelike tetrahedron.

\subsubsection{Spacelike triangle in spacelike tetrahedron}\label{Spacelike triangle in spacelike tetrahedron}

Firstly, let us consider a simple example: The bivector $\mathbf{X}_0=\hat{t}\wedge \hat{z}$, where $\hat{t}=(1,0,0,0)^T$ and $\hat{z}=(0,0,0,1)^T$, corresponds to a spacelike triangle lying in the $x$-$y$ plane. The spin-1/2 representation of ${\bf X}_0$ relates to the spinor $\xi_0=(1,0)^T$ by 
\be
X_0=\frac{\sigma^3}{2}=\left(\xi_0\otimes\xi_0^{\dagger}\right)-\frac{\mathbb{1}}{2}.\label{X0xi0}
\ee

Any ${\bf X}_{ef}$ of spacelike triangle in spacelike tetrahedron is given by
\be
{\bf X}_{ef}=\hat{t}\wedge \hat{n}_{ef}.
\ee
The 3d normal of the triangle $\hat{n}_{ef}$ is an SO(3) rotation of $\hat{z}$, while the SO(3) rotation leaves the 4d normal $\hat{t}$ invariant. Correspondingly, $X_{ef}$ relates to $X_0$ by the SU(2) adjoint transformation
\be 
 X_{ef}=S_{\Su} X_0 S_{\Su}^{-1} . \label{XefX0}
\ee 
Therefore, $X_{ef}$ can be written as
\be 
X_{ef}= \left(\xi_{ef}\otimes\xi_{ef}^{\dagger}\right)-\frac{\mathbb{1}}{2},\qquad \xi_{ef}=S_{\Su} \xi_0.\label{bivectorXsu2}
\ee 
The bivector with the above relation is always traceless, since $\langle \xi_{ef},\xi_{ef} \rangle =1$. 

We compute the $\Su$ matrix $S_{\Su}$, we solve the following equation which is equivalent to \eqref{XefX0}
\be 
S_{\Su}\sigma_3 S_{\Su}^{-1} &=& E \sigma_{1} + F \sigma_2+ G \sigma_3,\label{SsigS}
\ee 
where $\hat{n}_{ef}=(0, E,F,G)^T=(0,\langle\xi_{ef}|\vec{\sigma}|\xi_{ef}\rangle)$, or equivalently
\be 
E=\mathrm{Tr}(X_{ef}\sigma_1),\quad F = \mathrm{Tr}(X_{ef}\sigma_2),\quad G= \mathrm{Tr}(X_{ef}\sigma_3).
\ee
To solve \eqref{SsigS} for $S_{\Su}$, we firstly parametrize $S_{\Su}=\left(\begin{matrix}
	a_1+\imath a_2 & b_1+\imath b_2\\
	-b_1+\imath b_2 & a1 -\imath a_2
\end{matrix}\right)$ without imposing $\det(S_{\Su})=1$. Eq.\eqref{SsigS} is invariant under $S_{\Su}\to S_{\Su} e^{i\theta \sigma_3}$ for any $\theta\in[0,2\pi)$, so we can at most determine $S_{\Su}$ up to a phase. We use this freedom to fix $a_1>0$ and $a_2=0$. Then \eqref{SsigS} gives 3 independent equations determining uniquely $a_1,b_1,b_2$ after imposing  $\det(S_{\Su})=1$. Then, the SU(2) spinor $\xi_{ef}$ for the spacelike triangle is obtained by $\xi_{ef} = S_{\Su}\xi_0$.

\subsubsection{Spacelike triangle in timelike tetrahedron} 

Recall the example at the beginning of Section \ref{Spacelike triangle in spacelike tetrahedron}. We rename $X_0$ by $X_0^+$ and $\xi_0$ by $\xi_0^+$, and we rewrite \eqref{X0xi0} by
\be
X_0^+=\frac{\sigma^3}{2}=\sgn(\langle\xi_0^+,\xi_0^+\rangle)\left(\xi_{0}^+\otimes\xi_{0}^{+\dagger}\right)\sigma^3-\frac{\mathbb{1}}{2}.
\ee
where $\langle a,b\rangle = a^{\dagger}\sigma_3 b$ is the $\Suone$ invariant inner product. We check that the bivector $-\mathbf{X}_0=(-\hat{t})\wedge\hat{z}$ has the spin-1/2 representation 
\be
X_0^-=-\frac{\sigma^3}{2}=\sgn(\langle\xi_0^-,\xi_0^-\rangle)\left(\xi_{0}^-\otimes\xi_{0}^{-\dagger}\right)\sigma^3-\frac{\mathbb{1}}{2}.
\ee
where $\xi_0^-=(0,1)$

Any $\mathbf{X}_{ef}$ of spacelike triangle (in timelike tetrahedron) is given by 
\be
\mathbf{X}_{ef}^+=\hat{n}_{ef}^+\wedge \hat{z},\quad \text{or}\quad \mathbf{X}_{ef}^-=\hat{n}_{ef}^-\wedge \hat{z},
\ee
where $\hat{z}=N_{\rm ref}$ for the timelike tetrahedron, and $\hat{n}_{ef}^\pm$ is the future/past pointing timelike normal of $f$. $\hat{n}_{ef}^\pm$ relates to $\pm \hat{t}$ by an SU(1,1) transformation, while $\hat{z}$ is left invariant. Therefore $\mathbf{X}_{ef}^\pm$ relates to $\mathbf{X}_0^\pm$ by SU(1,1) adjoint transformation, and accordingly, in the spin-1/2 representation
\be
X_{ef}^\pm &=&S_{\Suone}X_0^\pm S_{\Suone}^{-1}
=\sgn\left(\langle\xi^\pm_{ef},\xi^\pm_{ef}\rangle\right) \left(\xi^\pm_{ef}\otimes\xi_{ef}^{\pm\dagger}\right)\sigma^3-\frac{\mathbb{1}}{2},\label{bivectorX}\\
&&\text{where}\qquad \xi_{ef}^\pm = S_{\Suone}\xi_0^\pm.
\ee 
We have used $ S_{\Suone}^\dagger\sigma^3 S_{\Suone}=\sigma^3$. Given the bivector $X^\pm_{ef}$, we want to solve the following equations for $S_{\Suone}$: 
\be 
&&S_{\Suone}\sigma_3 S_{\Suone}^{-1} = \pm \left[E^\pm \sigma^{3} + F^\pm (\imath\sigma^1)+ G ^\pm(\imath \sigma^2)\right],\\
&& {E^\pm=\mathrm{Tr}(X^\pm_{ef}\sigma^3),\quad F^\pm = \mathrm{Tr}(X^\pm_{ef}\left(-\imath \sigma^1\right)),\quad G^\pm = \mathrm{Tr}(X^\pm_{ef}\left(-\imath \sigma^2\right)).} \label{EFG}
\ee 
where $E^\pm,F^\pm,G^\pm$ relates to $\hat{n}_{ef}^\pm$ by $\hat{n}_{ef}^\pm=(E^\pm,-G^\pm,F^\pm,0)^T$. The procedure of solving \eqref{EFG} is very similar to solving \eqref{SsigS} for $S_{\Su}$: we parametrize $S_{\Suone}=\left(\begin{matrix}
	a_1+\imath a_2 & b_1+\imath b_2\\
	b_1-\imath b_2 & a1 -\imath a_2
\end{matrix}\right)$ without imposing $\det(S_{\Suone})=1$. Eq.\eqref{EFG} is invariant under $S_{\Suone}\to S_{\Suone} e^{i\theta \sigma_3}$, where $e^{i\theta \sigma_3}\in\Suone$ for any $\theta\in[0,2\pi)$. We use this freedom to fix $a_1>0$ and $a_2=0$. Then \eqref{EFG} gives 3 independent equations determining uniquely $a_1,b_1,b_2$ after imposing  $\det(S_{\Suone})=1$. The SU(1,1) spinor $\xi^\pm_{ef}$ for the spacelike triangle is obtained by $\xi_{ef}^\pm = S_{\Suone}\xi_0^\pm$.

Generally a spinor $\xi\in \C^2$ determines a 4d null vector 
$V^I(\xi)= \xi^\dagger\sigma^I \xi = (\iota_{\Suone}(\xi), V^3(\xi))^T$, where $I=0,1,2,3$. One can check that, for a $\Suone$ spinor $\xi^\pm_{ef}$, the null vector are always future pointing with 
\be 
V(\xi_{ef}^+)= (\iota_{\Suone}(\xi_{ef}^+), 1)^T, \quad V(\xi_{ef}^-)= (\iota_{\Suone}(\xi_{ef}^-), -1)^T,
\ee  
especially $V(\xi^+_0)= (1,0,0,1)^T$ and $V(\xi_0^-)= (1,0,0,-1)^T$. The bivector can be written as
\be 
\mathbf{X}_{ef}^\pm = \pm V(\xi_{ef}^\pm) \wedge \hat{z},
\ee 
and we have $n_{ef}^\pm = \left(\pm \iota_{\Suone},0\right)^T $.

\subsubsection{Timelike triangle in timelike tetrahedron}

The bivector $\mathbf{X}_{0}'=\hat{y}\wedge \hat{z}$ has the spin-1/2 representation is
\be
X_{0}'=\frac{\imath}{2}\sigma^1=\imath \left(l_{0}^+\otimes(l_{0}^-)^{\dagger}\right)\sigma^3-\frac{\imath}{2}\mathbb{1},\qquad l^{+}_0=\frac{1}{\sqrt{2}} \left(\begin{matrix}
	1\\
	1
\end{matrix}\right), \quad l^{-}_0=\frac{1}{\sqrt{2}} \left(\begin{matrix}
	1\\
	-1
\end{matrix}\right).
\ee
where $\langle\ ,\ \rangle$ is the SU(1,1) invariant inner product. A generic bivector $\mathbf{X}_{ef}$ of timelike triangle $f$ is 
\be
\mathbf{X}_{ef}=\hat{n}_{ef}\wedge\hat{z}
\ee
with spacelike $\hat{n}_{ef}$ and relates to $\mathbf{X}_{0}'$ by the SU(1,1) adjoint transformation. Therefore we obtain
\be 
X_{ef}= S_{\Suone} X_0' S_{\Suone}^{-1} =\imath \left(l_{ef}^+\otimes(l_{ef}^-)^{\dagger}\right)\sigma^3-\frac{\imath}{2}\mathbb{1},\qquad l_{ef}^\pm = S_{\Suone} l^\pm_0.\label{bivectort}
\ee 
Given the bivector $X_{ef}$ for the timelike triangle, we solve for $S_{\Suone}$ from
\be 
&&S_{\Suone}\left(\imath \sigma^1\right) S_{\Suone}^{-1} = E \sigma^{3} + F(\imath\sigma^1)+ G (\imath \sigma_2), \label{EFGtime}\\
&&{ E=\mathrm{Tr}(X_{ef}\sigma^3),\quad F = \mathrm{Tr}(X_{ef}\left(-\imath \sigma^1\right)),\quad G =\mathrm{Tr}(X_{ef}\left(-\imath \sigma^2\right)).}
\ee 
The normal vector $\hat{n}_{ef}$ relates to $E,F,G$ by $\hat{n}_{ef}=(E,-G,F,0)^T$. The equation \eqref{EFGtime} is left invariant by $S_{\Suone}\to - S_{\Suone} $ and $S_{\Suone}\to S_{\Suone}e^{\theta \sigma^1}$ where $e^{\theta \sigma^1}\in\Suone$ for any $\theta\in\R$.
We parametrize $S_{\Suone}=\left(\begin{matrix}
	a_1+\imath a_2 & b_1+\imath b_2\\
	b_1-\imath b_2 & a1 -\imath a_2
\end{matrix}\right)$ without imposing $\det(S_{\Suone})=1$. We use the symmetries to fix $a_1>0$ and $b_1=0$. Then \eqref{EFGtime} gives 3 independent equations determining uniquely $a_1,a_2,b_2$ after imposing  $\det(S_{\Suone})=1$. The SU(1,1) spinor $\xi^\pm_{ef}$ for the spacelike triangle is obtained by $l_{ef}^\pm = S_{\Suone} l^\pm_0$.

The spinors $l_{ef}^\pm\in\C^2$ determines the complex null vector $U^I(l_{ef}^+,l_{ef}^-)=-\imath l_{ef}^{+\dagger}\sigma^I l^-_{ef}$, $I=0,1,2,3$ \footnote{A null tetrad is given by $l_{ef}^{+\dagger}\sigma^I l^+_{ef},\ l_{ef}^{-\dagger}\sigma^I l^-_{ef},\ l_{ef}^{+\dagger}\sigma^I l^-_{ef},\ l_{ef}^{-\dagger}\sigma^I l^+_{ef}$.}
\be
U(l_{ef}^+,l_{ef}^-)=\lt(\iota(l_{ef}^+,l_{ef}^-),\imath\rt)^T
\ee
The null vector relates $\hat{n}_{ef}$ and $\mathbf{X}_{ef}$ by
\be
\hat{n}_{ef}=\lt(\iota(l_{ef}^+,l_{ef}^-),0\rt)^T,\qquad \mathbf{X}_{ef}=U(l_{ef}^+,l_{ef}^-)\wedge\hat{z}.
\ee

\vspace{5mm}

\begin{figure}[h]
    \centering
    \includegraphics[scale=0.35]{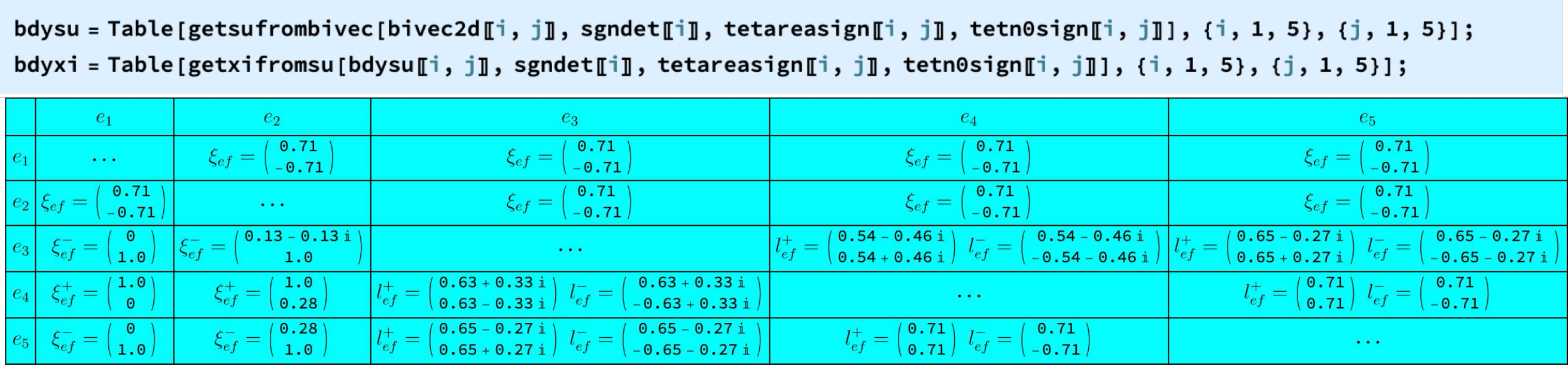}
    \caption{Computation of $S_{\text{SU(2)}}$, $S_{\text{SU(1,1)}}$ matrices, spinors, and numerical results of spinors: Each cell of the table represents the spinor for the face shared by the tetrahedron labelling the row and the tetrahedron labelling the column. The spinors in each row are for the same tetrahedron. $\xi_{ef}$ denotes the spinor for the spacelike face in the spacelike tetrahedron, $\xi^{\pm}_{ef}$ denotes the future/past-pointing spinor for the spacelike face in the timelike tetrahedron, and $l^\pm_{ef}$ denotes the spinor for the timelike face in the timelike tetrahedron.}
    \label{fig:Spinors}
\end{figure}

Based on the algorithm provided above, in FIG. \ref{fig:Spinors}, we demonstrate the application of the function \textbf{getsufrombivec} to compute the SU(2) and SU(1,1) matrices. This computation involves using the input parameters, including the bivectors \textbf{bivec2d} from FIG. \ref{fig:Bivec}, the types of tetrahedra \textbf{sgndet}, the types of triangles \textbf{tetareasign}, and \textbf{tetn0sign}, which can take values of $+1$ or $-1$ depending on the orientation of the timelike normal (either future or past pointing). The resulting numerical values of the SU(2) and SU(1,1) matrices are stored in \textbf{bdysu}. Additionally, we perform a numerical computation of the spinors for all triangles using the function \textbf{getxifromsu} with \textbf{bdysu} as the input. The spinors specify the boundary state for the 4-simplex amplitude. The numerical results are shown in FIG. \ref{fig:Spinors}.

\subsection{3d face normals $n_{ef}$}

We would like to compute the outgoing 3d face normals $n_{ef}$ of each tetrahedron $e$. Generally, $n_{ef}$ is different from $\hat{n}_{ef}$ up to $\pm$ sign and removing a trivial component. The 4-simplex amplitude depends on the spinors that are computed in the last subsection, whereas the outgoing normals $n_{ef}$ are not immediately useful at the level of a single 4-simplex, but they will be useful when discussing gluing 4-simplices. 

\begin{figure}[h]
    \centering
    \caption{Computation of $P^{(3)}_{e,i}$}
    \includegraphics[scale=0.55]{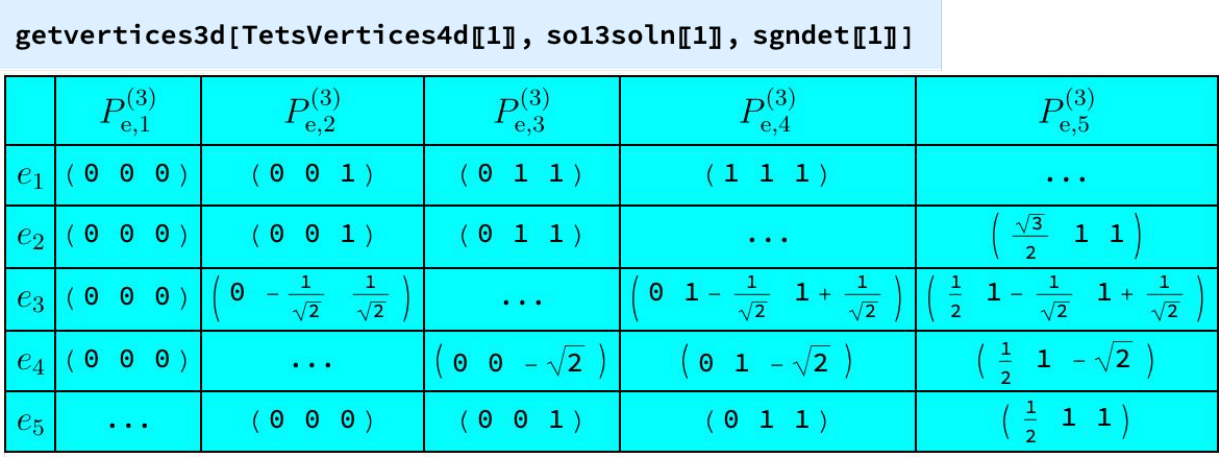}
    \label{fig:vertices3d}
\end{figure}

To obtain the 3d face normals for each face, we need to embed each tetrahedron in three dimensions. In each tetrahedron $e$, the 3d coordinates $P^{(3)}_{e,i}$ of each vertex $i$ belonging to $e$ can be obtained by dropping the first or last component of 
\be 
P_{e,i}^{(4)} = \Lambda^{-1}_{ve} P_{i}, 
\ee 
where $P_{i}$ is the 4D coordinates of each vertex as defined in (\ref{vertices}). In FIG. \ref{fig:vertices3d}, we demonstrate the function {\bf getvertices3d} for computation of $P^{(3)}_{e, i}$ using the input of 4d vertices \textbf{TetsVertices4d}, the type of the tetrahedron \textbf{sgndet}, and \textbf{so13soln} containing the data  $\L_{ve}\in\,$SO(1,3) for each tetrahedron, while also presenting the numerical results of the corresponding 3d coordinates.

\begin{figure}[ht]
    \centering
    \includegraphics[scale=0.5]{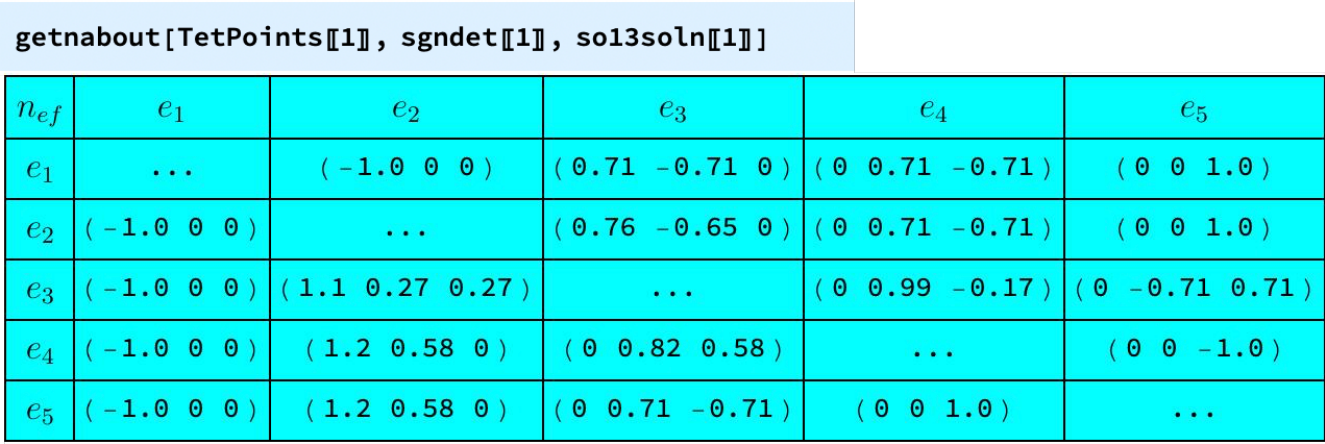}
    \caption{Numerical results of outgoing face normals $n_{ef}$: Each cell of the table is the 3d face normals for the face shared by the line tetrahedron $e_i$ and the column tetrahedron $e_j$.}
    \label{fig:3dnormals}
\end{figure}

With the given 3d coordinates, we can compute the 3d edge vectors $P^{(3)}_{e, ij} = P^{(3)}_{e,i} - P^{(3)}_{e,j}$ connecting each pair of 3d vertices. For each face of a tetrahedron, the 3d face normal $n_{ef}$ is perpendicular to all 3d edges of the face, which means that 
\be
n_{ef}\cdot\eta^{(3)}\cdot P^{(3)}_{e, ij} = 0.\label{orthogonal}
\ee
Here, the metric $\eta^{(3)}=\mathrm{diag}(1,1,1)$ when the tetrahedron is spacelike, and $\eta^{(3)}=\mathrm{diag}(-1,1,1)$ when the tetrahedron is timelike. For each face, there are only 2 independent edge vectors $P^{(3)}_{e, ij}$ leading to 2 linear equations that determine $n_{ef}$ up to scaling. Subsequently, we normalize each vector to obtain a unit face normal vector. However, this unit face normal vector can be either incoming or outgoing to the tetrahedron. The procedure of determining outgoing normals $n_{ef}$ is similar to the procedure of determining the outgoing 4d normals of tetrahedra in 4-simplex: the tetrahedron defined by the coordinates $P_{e,i}^{(3)}$ is a convex hull:
\be 
T = \left\{\sum_{i=1}^4 t_i P^{(3)}_{e,i}; \quad t_i\geq 0, \quad \sum_{i=1}^4 t_i=1 \right\}.
\ee
If $n_{vf}$ represents an outgoing normal originating from the barycenter $\fp_f$ of the face $f$, then $\fp_f+\epsilon n_{vf}$ for any $\epsilon>0$ lies outside $T$. With these constraints, we can proceed to select the corresponding outgoing 3d face normals. In FIG.\ref{fig:3dnormals}, we show the corresponding normalized outgoing 3d face normals for this given 4-simplex. 

\subsection{Triangle orientations $\kappa_{ef}$}

\begin{figure}[ht]
    \centering
    \includegraphics[scale=0.3]{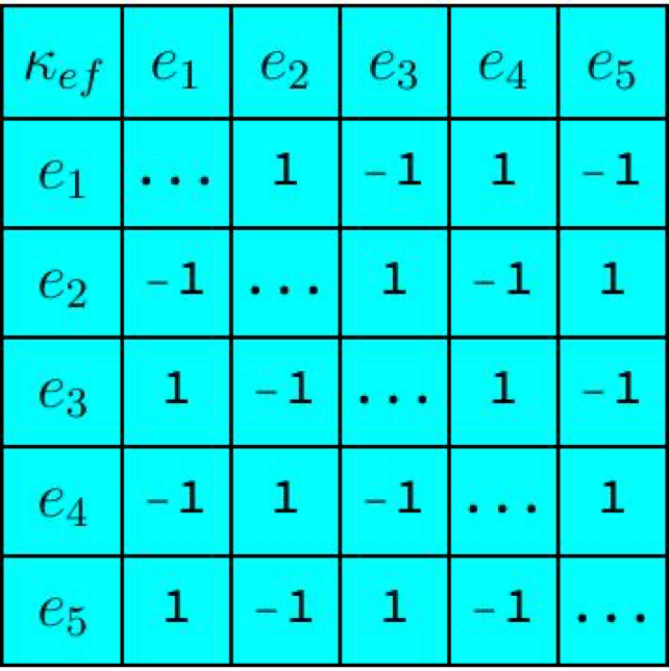}
    \caption{Orientation of triangles in 4-simplex.}
    \label{fig:kappa1}
\end{figure}

To define the orientations of the triangles in a 4-simplex, we assign $\kappa_{ef}=\pm 1$ such that it satisfies the relation $\kappa_{ef}=-\kappa_{e'f}$ for $e,e'$ sharing $f$. The set of $\kappa_{ef}$ is determined up to a global sign and satisfies the closure condition
\be
\sum_{i=1}^4\kappa_{ef_i} X_{ef_i} A_{f_i} = 0.
\ee
Here, we adapt the orientation convention outlined in FIG. \ref{fig:kappa1}.

\subsection{Spinors $z_{vf}$}\label{sec:zvf}

\begin{figure}[ht]
    \centering
    \caption{Computation of $z_{vf}$}
    \includegraphics[scale=0.4]{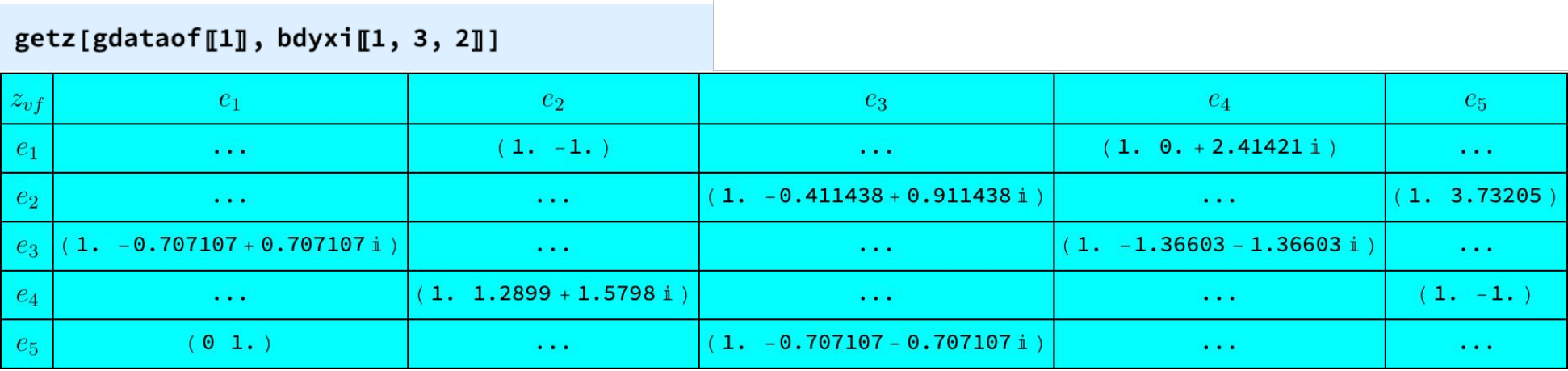}
    \label{fig:zvf}
\end{figure}

The spinor $z_{vf}$ is fixed up to a complex scaling by $g_{ve}$ and spinors:
\be
z_{vf}\propto_\C \begin{cases} (\mathring{\fg}_{ve}^{\text{T}})^{-1} \xi_{ef},\quad \text{spacelike face in spacelike tetrahedron}\\
(\mathring{\fg}_{ve}^{\text{T}})^{-1} \xi^\pm_{ef},\quad \text{spacelike face in timelike tetrahedron}\\
(\mathring{\fg}_{ve}^{\text{T}})^{-1} l^-_{ef},\quad \text{timelike face in timelike tetrahedron}
\end{cases}
\ee
Complex scaling $z_{vf}$ is a gauge transformation of the spinfoam action $S_f$ in (\ref{zvf}). To remove this scaling gauge freedom, we set the first component of $z_{vf}$ to 1 (or the second component to 1 if the first element is 0) and choose its form to be $z_{vf}=(1,\alpha_{vf})^\text{T}$ (or $z_{vf}=(0,1)^\text{T}$). In FIG. \ref{fig:zvf}, we demonstrate the function \textbf{getz} computing $z_{vf}$ using the input of the $\mathrm{SL}(2,\mathbb{C})$ group elements \textbf{gdataof}, and the spinors $\textbf{bdyxi}$, along with presenting the numerical results. 

\subsection{4-simplex action}

\begin{figure}
    \centering
    \caption{Computation of action at the critical point}
    \includegraphics[scale=0.5]{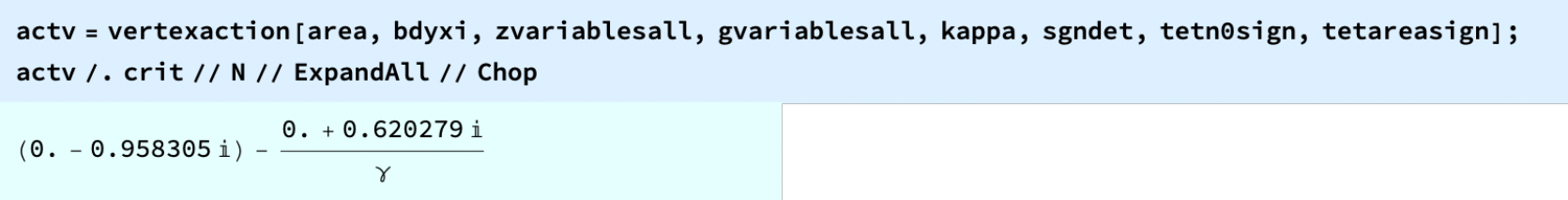}
    \label{fig:action}
\end{figure}

\begin{figure}
    \centering
    \caption{Computation of the derivative of action at the critical point}
    \includegraphics[scale=0.5]{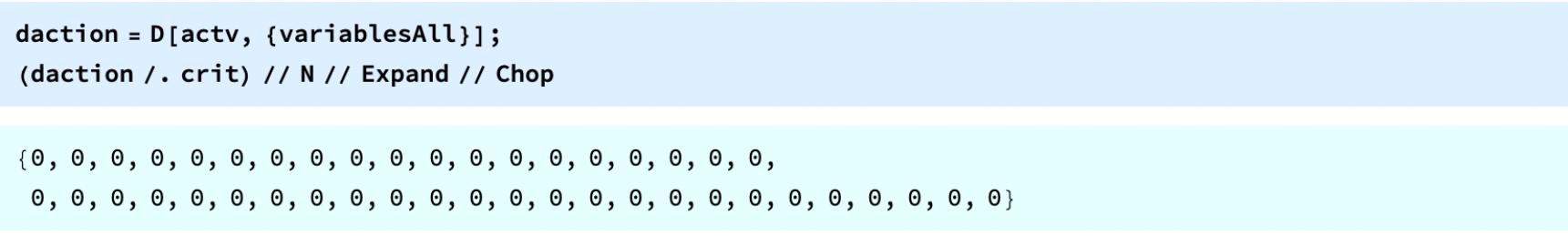}
    \label{fig:eom}
\end{figure}

We denote by $\{\mathring{\fg}_{ve}, \mathring{z}_{vf}\}$ the critical point constructed above. In order to check it is indeed the critical point, we parametrize $\fg_{ve}$ and $z_{vf}$ in a neighborhood of the critical point within the integration domain of (\ref{SFamplitude}). The parametrization of $\Slc$ group element is given by
\be
\fg_{ve}\left(x^1_{ve},y^1_{ve},x^2_{ve},y^2_{ve},x^3_{ve},y^3_{ve}\right)=\left(\begin{matrix}1+x_{ve}^1+\imath y_{ve}^2&x_{ve}^2+\imath y_{ve}^2\\
x_{ve}^3+ \imath y^3_{ve}&*\end{matrix}\right),
\label{ge}
\ee 
where $x_{ve}^i$ and $y_{ve}^i$ (for $e\neq e_1$, $i=1,2,3$) are real, and the entry $*$ is determined by $\det(\fg_{ve})=1$. $\fg_{ve_1}$ is fixed to be a constant matrix\footnote{We fix this constant matrix to be the same as $\mathring{\fg}_{ve_1}$ at the critical point.}, leaving us with 24 real variables, $x_{ve}^i$ and $y_{ve}^i$ (for $e=e_2,e_3,e_4,e_5$), to parameterize all $g_{ve}$. The corresponding critical point data $(\mathring{x}_{ve}^i, \mathring{y}_{ve}^i)$ can be determined by comparing Eq.(\ref{ge}) with Table \ref{fig:sl2csoln}. The spinor $z_{vf}\in \mathbb{CP}^{1}$ is parametrized as
\be 
z_{vf} = \left(\begin{matrix}
    1\\ x_{vf}+\imath y_{vf}
\end{matrix}\right), \label{zvfpara}
\ee 
where $x_{vf}, y_{vf} \in \mathbb{R}$. Each face $f$ has 2 real variables $x_{vf}$ and $y_{vf}$, so we need a total of 20 real variables to parametrize all $z_{vf}$. Using the boundary data $(j_{f}, \xi_{ef}, \xi^{\pm}_{ef}, l^+_{ef})$, the critical point data $(\mathring{x}_{ve}^i, \mathring{y}_{ve}^i, \mathring{x}_{vf}, \mathring{y}_{vf})$, and the parameterization described earlier, we apply Eq.(\ref{SF}) to obtain the 4-simplex action. The resulting action $S[\vec{\mathbf{x}}]=S[x_{ve}^i, y_{ve}^i, x_{vf}, y_{vf}]$ has 44 real parameters. In our code, the real critical points $(\mathring{\fg}_{ve}, \mathring{z}_{vf})$ are stored in $\textbf{crit}$, with the input of boundary data $\textbf{area}$ and $\textbf{bdyxi}$, the variables $\textbf{zvariablesall}$ and $\textbf{gvariablesall}$, orientation $\textbf{kappa}$, and the types of tetrahedra and faces $\textbf{sgndet}$, $\textbf{tetn0sign}$, $\textbf{tetareasign}$. We compute the action of the vertex $\textbf{actv}$ in FIG \ref{fig:action}, which is a function in terms of 44 real variables. Then, we can evaluate $\textbf{actv}$ at the real critical points $\textbf{crit}$, and the result is shown in FIG \ref{fig:action}. Then it is straight-forward to compute the derivative of the action and show that the derivatives with respect to all $x_{ve}^i, y_{ve}^i, x_{vf}, y_{vf}$ vanish at the critical point (see FIG. \ref{fig:eom}).

\subsection{Parity transform}

\begin{figure}[ht]
    \centering
    \caption{Computation of $\Slc$ solution for the orientation $s_v=+$}
    \includegraphics[scale=0.6]{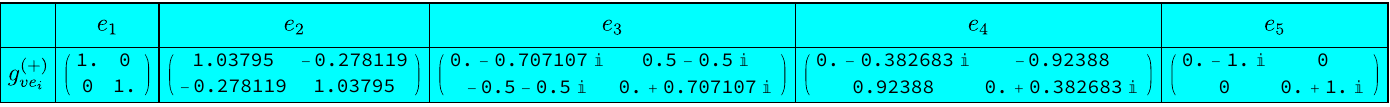}
    \label{fig:sl2csolnminus}
\end{figure}

\begin{figure}[h]
    \centering
    \caption{Computation of $\mathring{z}^{(+)}_{vf}$ for the orientation $s_v=+$}
    \includegraphics[scale=0.6]{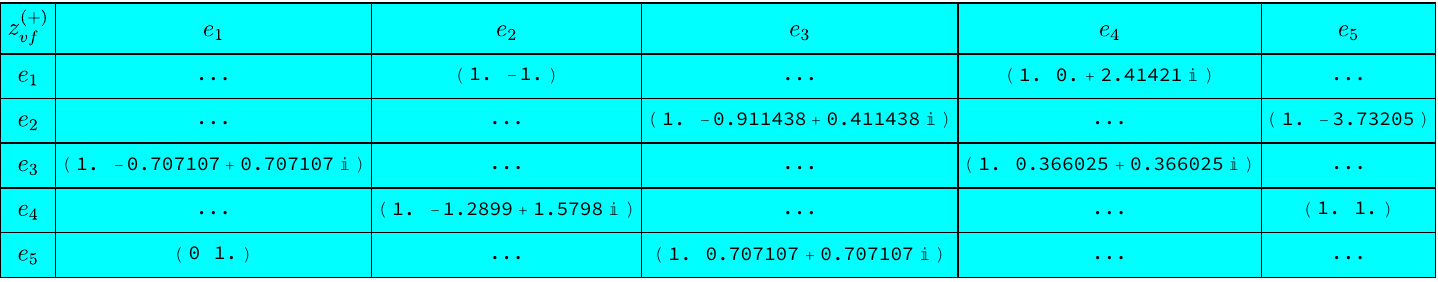}
    \label{fig:zvfminus}
\end{figure}

So far we have considered the real critical point $\mathring{x}=\{\mathring{\fg}_{ve},\mathring{z}_{vf}\}$. Given the boundary data $\mathring{r}=\{\xi_{ef},\xi^\pm_{ef},l^+_{ef},j_{f}\}$ constructed from the 4-simplex geometry, there are exactly 2 real critical points $\mathring{x}$ and $\mathring{x}'$. They correspond two different orientations of the 4-simplex, and we denote the orientation of $\mathring{x}$ by $s_v=-$ and the orientation of $\mathring{x}'$ by $s_v=+$. The $\Slc$ solution $g_{ve}^{(s_v)}$ for different orientations have the following relation \cite{Han:2021rjo}
\be 
g_{ve}^{(+)} = \begin{cases} \left(g^{(-)\dagger}_{ve}\right)^{-1},\quad &e\text{ is spacelike tetrahedron}\\
\left(g^{(-)\dagger}_{ve}\right)^{-1}.\left(\imath\sigma_3\right),\quad &e\text{ is timelike tetrahedron}
\end{cases} \label{gveminus}
\ee 
here, the result of $g_{ve}^{(-)}$ is shown in FIG \ref{fig:sl2csoln}. The result of the $\Slc$ solution for the orientation $s_v=+$ is in FIG. \ref{fig:sl2csolnminus}. 
Then, one can compute the $\Slc$ group element $\mathring{\fg}_{ve}^{(+)}$ in the spinfoam action (\ref{SF}) for the orientation $s_v=+$ using (\ref{gvegve}). After that, following the procedure to compute spinors $z_{vf}$ in Sec. \ref{sec:zvf}, we can compute the corresponding solutions of spinors $\mathring{z}^{(+)}_{vf}$ by using the function \textbf{getz} with the input of $\mathring{\fg}^{(+)}_{ve}$ and the boundary data $\xi_{ef}$ in FIG \ref{fig:Spinors}. The result of $\mathring{\fg}^{(+)}_{ve}$ is presented in FIG \ref{fig:zvfminus}. 
Finally, we can compute the 4-simplex action at the real critical point $\{\mathring{\fg}^{(+)}_{ve},\mathring{z}^{(+)}_{vf}\}$ for the orientation $s_v=+$ using the function \textbf{vertexaction}. The numerical result of $S^{(+)}_v$ is shown in FIG \ref{fig:actionsvminus}. 
\begin{figure}[h]
    \centering
    \caption{Computation of action at the real critical point $\{\mathring{\fg}^{(+)}_{ve},\mathring{z}^{(+)}_{vf}\}$}
    \includegraphics[scale=0.5]{figures/actionsvplus.pdf}
    \label{fig:actionsvminus}
\end{figure}
However, the 4-simplex amplitude involves an arbitrary phase due to the phase ambiguity of $\xi_{ef},l^+_{ef}$. The overall phase can be determined by 
\be 
\phi_v = \frac{1}{2}\left(S_v^{(+)}+S_v^{(-)}\right).
\ee 
The phase difference between two critical point is 
\be
\Delta S_v=S_v^{(+)}-S_v^{(-)}.
\ee
The asymptotics of the 4-simplex amplitude is given by 
\be
e^{i\phi_v}\lt(\cn_+ e^{\Delta S_v/2}+\cn_- e^{-\Delta S_v/2}\rt)\lt[1+O(1/\l)\rt].
\ee

We check numerically that the phase difference $\Delta S_v$ relates to the Regge action $S_{\rm Regge}$ by \footnote{When generalize to simplicial complex, this difference only happens for boundary $f$ thus only contributes to the boundary term, while the bulk terms coincide with the Regge action \cite{Han:2024ydv,Liu:2018gfc}. When the timelike triangle is absent, this difference turns out to be $2\pi i\Z$ thus can be absorbed into the overall phase \cite{Kaminski:2017eew}.} 
\be 
\Delta S_v - 2 \imath S_{\rm Regge} = \sum_{f\in t\text{-}s} (\pm\imath {\pi})j_f + {\sum_{f\in t\text{-}t}} \left(2\imath \pi\text{ or } 0\right)j_f, \label{dsdreggev}
\ee 
where $f\in t\text{-}s$ denotes the spacelike face in timelike tetrahedron, and $f\in t\text{-}s$ denotes the timelike face. The Regge action is defined as
\be 
S_{\rm Regge} =\sum_{1\leq i<j\leq5} \left| A_{e_ie_j}\right| \tilde{\theta}_{e_ie_j},
\ee 
where $\left| A_{e_ie_j}\right|$ is the absolute value of the area of the face between tetrahedra $e_i$ and $e_j$ (as illustrated in Figure \ref{fig:areas}). The dihedral angle between tetrahedra $e_i$ and $e_j$ is denoted by $\tilde{\theta}_{e_ie_j}$, where $\tilde{\theta}_{e_ie_j}= \theta_{e_ie_j}$ for spacelike faces and $\tilde{\theta}_{e_ie_j}= \pi-\theta_{e_ie_j}$ for timelike faces, and $\theta_{e_ie_j}$ is given in Figure \ref{fig:areas}. The result in FIG.  \ref{fig:dsdregge} confirms (\ref{dsdreggev}). 
\begin{figure}
    \centering
    \caption{The numerical result of (\ref{dsdreggev}) includes spins $j_f\in\frac{\mathbb{N}}{2}$ within the 4-simplex action. Here, we do not evaluate the spins (e.g. ${\rm j\$1\$3}$ is the spin at the triangle shared by $e_1$ and $e_3$) and maintain the form to make a clear comparison with (\ref{dsdreggev}). Among all spins, $\text{j$\$$1$\$$3},\text{j$\$$1$\$$4},\text{j$\$$1$\$$5},\text{j$\$$2$\$$3},\text{j$\$$2$\$$4}$ and $\text{j$\$$2$\$$5}$ correspond to the spacelike triangles in timelike tetrahedra ($s\text{-}t$), while the spins $\text{j$\$$3$\$$4}$ and $\text{j$\$$4$\$$5}$ correspond to the timelike triangle ($t\text{-}t$). }
    \includegraphics[scale=0.55]{figures/dsdregge.pdf}
    \label{fig:dsdregge}
\end{figure}

\section{Spinfoam on simplicial complex}\label{SimplicialComp}

In this section, we generalize our computation to real critical points of the spinfoam amplitude on simplicial complex. For the purpose of demonstrating the algorithm, we consider the example of the complex containing only two 4-simplices, $v_1$ and $v_2$, sharing a common bulk tetrahedron. The idea is to independently construct the boundary data and critical points for each $v_i$ and then match them at the shared tetrahedra. For clarity, we will discuss separately two cases, where the internal tetrahedron is spacelike and timelike.

\subsection{Internal tetrahedron is spacelike}

\begin{figure}[ht]
	\centering
	\includegraphics[scale=0.12]{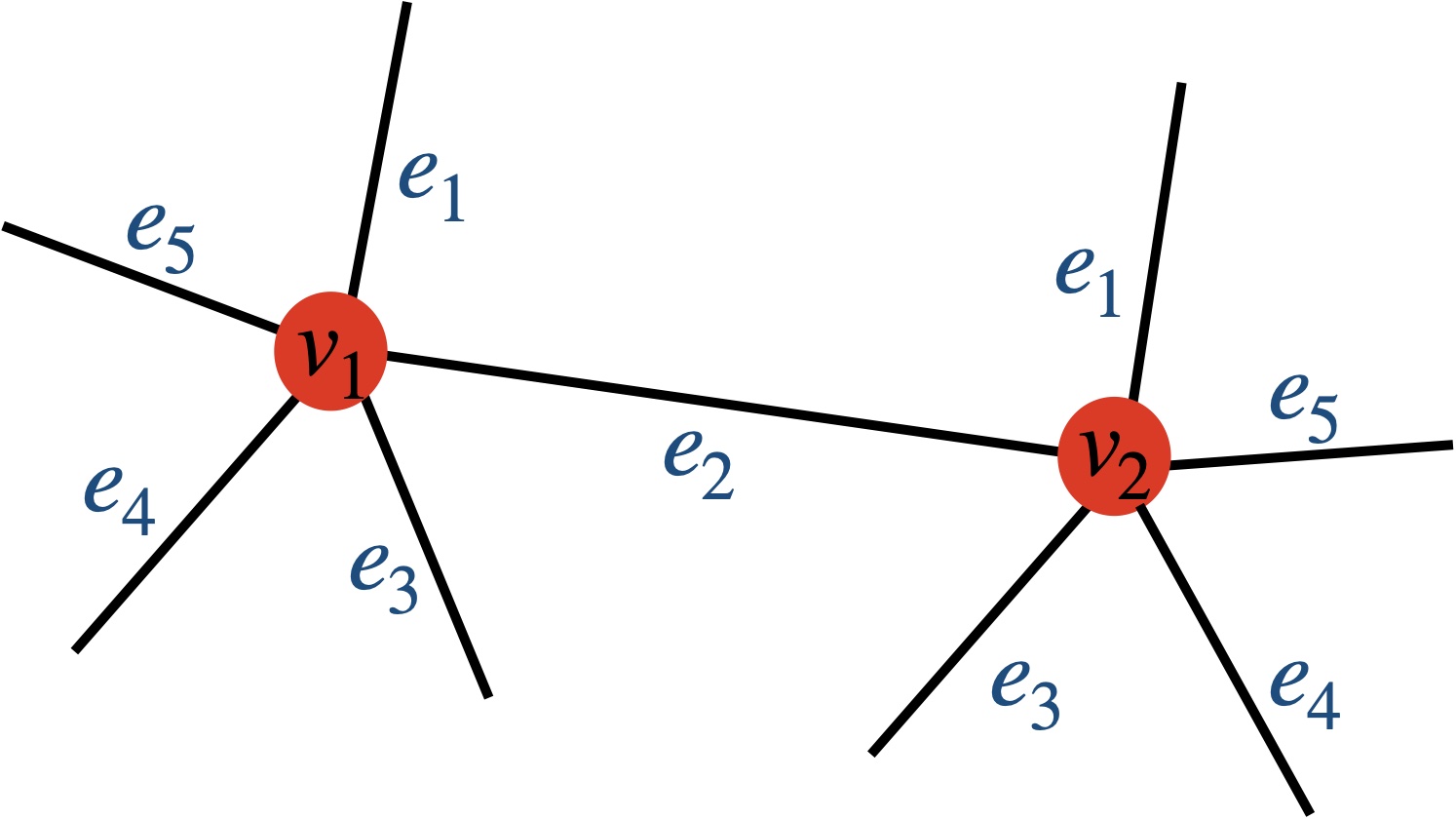}
 \caption{Dual diagram for two 4-simplices share one spacelike bulk tetrahedron} \label{4simplex2}
\end{figure}

We use the same coordinates as in Eq. (\ref{vertices}) for the first 4-simplex, $v_1$. For the second 4-simplex, $v_2$, we use the following coordinates for its veritces:
\be
P_1 = (0,0,0,0),\quad
P_2 = (0,0,0,1),\quad
P_3 = (0,0,1,1),\quad 
P_{4'} = (\frac{1}{2},0,1,1),\quad
P_5 = (\frac{1}{2},1,1,1).\label{vertices2}
\ee 
The only difference between (\ref{vertices}) and (\ref{vertices2}) is the coordinates of $P_4$ and $P_{4'}$. The tetrahedra in this simplicial complex are given by
\be 
\{v_1, e_i\}&=&\{\{1,2,3,4\},\{1,2,3,5\},\{1,2,4,5\},\{1,3,4,5\},\{2,3,4,5\}\}_{v_1},\\
\{v_2, e_i\}&=&\{\{1,2,3,4'\},\{1,2,3,5\},\{1,2,4',5\},\{1,3,4',5\},\{2,3,4',5\}\}_{v_2}. 
\ee 
The internal tetrahedron is $e_{2}=\{1,2,3,5\}$, see FIG. \ref{4simplex2}. We repeat the steps in section \ref{v1data} with the coordinates in (\ref{vertices2}) to obtain the corresponding boundary data $(j_{f}, \xi_{ef}, \xi^{\pm}_{ef}, l^+_{ef})$ and critical point $\{\mathring{\fg}_{v_2e},\mathring{z}_{v_2f}\}$ for the 4-simplex $v_2$.  

\subsubsection{Match face normals}

To glue two 4-simplices, the data $(j_f, \xi_{e_2f})$ of the shared tetrahedron $e_2$ must match between two 4-simplex amplitude on $v_1$ and $v_2$. The areas $j_f$ always match by construction, but matching $\xi_{e_2f}$ is nontrivial, since the boundary data of $v_1,v_2$ are constructed independently. We need to check whether the resulting $\xi_{e_2f}$ in $v_1$ and $v_2$ match. If not, we consistently use the $\xi_{e_2f}$ in $v_1$ as a reference to adjust the corresponding value of $\xi_{e_2f}$ in $v_2$. The following steps are performed to align the $\xi_{e_2f}$ values:

\begin{itemize}
    \item We denote by the 3d outgoing normals by $n_{e_2 f}(v_1)$ and $n_{e_2 f}(v_2)$ for $e_2$ in $v_1$ and $v_2$ respectively. We should match $n_{e_2 f}(v_1)$ and $n_{e_2 f}(v_2)$ for all 4 faces. Indeed, we can find an $\text{SO}(3)$ matrix $O_{e_2}$ to rotate $n_{e_2 f}(v_1)$ to $n_{e_2 f}(v_2)$ by solving the linear equations $n_{e_2f}(v)=O_{e_2}n_{e_2f}(v_2)$, $f=1,\cdots,4$, with the condition $\det O_{e_2}=1$. In FIG. \ref{fig:so3}, we demonstrate the function \textbf{SO3d} computating $O_{e_2}$ using the three inputs: the data of $n_{e_2f}(v_1)$, \textbf{nabref}, the data of $n_{e_2f}(v_1)$, \textbf{nabchange}, and the type of the tetrahedron $e_2$, $\textbf{sgndet[[1,2]]}$. 
    \begin{figure}[h]
        \centering
        \includegraphics[scale=0.6]{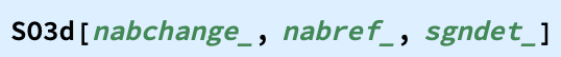}
        \caption{The computation of the $\mathrm{SO(3)}$ matrix to match face normals.}
        \label{fig:so3}
    \end{figure}
    
    \item The 3d $\text{SO}(3)$ matrix can be embedded into a 4d $\text{SO}(1, 3)$ matrix $\tilde{O}_{e_2} = \left(\begin{array}{*{4}{c}}
    1 & 0 & 0 & 0\\
    0 & & &  \\
    0 & & & \\
    0 & \multicolumn{3}{c}{\raisebox{\dimexpr\normalbaselineskip+.3\ht\strutbox-.3\height}[0.5pt][0.5pt]{$O_{e_2}$}} \\
    \end{array}\right)$. We then define the new $\text{SO}(1, 3)$ transformation $\tilde{\Lambda}_{v_2 e_2}=\tilde{O}_{ef}\Lambda_{v_2 e_2}$.

    \item We lift $\tilde{\Lambda}_{v_2 e_2}$ to $\Slc$ solutions $\tilde{g}_{v_2e_2}$ by solving the following equation (the convention is consistent with \eqref{gHgH} - \eqref{Lambdagg})
    \be
   (\tilde{\Lambda}_{v_2e_2}^{-1})^I_{ \ J} = \frac{1}{2} \Tr\left(\tilde{g}_{v_2e_2}\bar{\sigma}^I\tilde{g}_{v_2e_2}^\dagger\bar{\sigma}^J\right),\quad \bar{\sigma}^I = (\mathbb{1}_{2\times2}, -\vec{\sigma}). \label{getsl2c}
    \ee
    with the condition that $\det \tilde{g}_{v_2e_2}=1$. The sign of $\tilde{g}_{v_2e_2}$ is fixed arbitrarily, since it is a gauge freedom. We modify $\mathring{\fg}_{v_2e_2}$ to $\mathring{\fg}_{v_2e_2}=(\tilde{g}_{v_2e_2}^T)^{-1}$.
    
    \item We obtain 
    \be 
    \delta g_{v_2e_2}=\tilde{g}_{v_2e_2}
    \ee
    as the change from the previous solution $g_{v_2e_2}$ to the new solution $\tilde{g}_{v_2e_2}$.

    \item We can obtain the corresponding bivector $\tilde{X}_{e_2f}$ in $v_2$ by applying the following transformation:
    \be
    \tilde{X}_{e_2f}= (\delta g_{v_2 e_2})^{-1}X_{e_2 f}\delta g_{v_2e_2}.
    \ee
    \item Once we have $\tilde{X}_{e_2f}$ in $v_2$, we can use computation in Section \ref{xicomputation} to compute the corresponding $(\tilde{S}_{\Su})_{e_2}$ and $\tilde{\xi}_{e_2f}$ in $v_2$. At this point, $\tilde{\xi}_{e_2f}$ in the 4-simplex $v_2$ is the same as $\xi_{e_2f}$ in the 4-simplex $v_1$. 
\end{itemize}

\subsubsection{$\Su$ gauge fixing}

Due to the gauge freedom in (\ref{gaugesu2}) for the internal spacelike tetrahedron $e_2$, we fix $g_{v_1e_2}$ to be upper triangular matrix and adapt the following parametrization
\be
{g}_{v_1e_2}= \left(\begin{array}{cc}
	1+\frac{x_{v_1e_2}^{1}}{\sqrt{2}} & \frac{x_{v_1e_2}^{2}+\imath y_{v_1e_2}^{2}}{\sqrt{2}} \\
	0 & *
\end{array}\right), \label{gvenew}
\ee 
where the entry $*$ is determined by $\det(g_{ve})=1$. Any $\Slc$ group element $g$ can be written as $g=kh$ with $h\in \Su$ and $k$ upper triangular. The decomposition is explicitly given by 
\be 
\left(\begin{array}{ll}
a & c \\
b & d
\end{array}\right)=\left(\begin{array}{cc}
\lambda^{-1} & \mu \\
0 & \lambda
\end{array}\right)\left(\begin{array}{cc}
\bar{v} & -\bar{u} \\
u & v
\end{array}\right)\label{decomp}
\ee 
with
\be 
\lambda=\left(|b|^2+|d|^2\right)^{1 / 2},\quad \mu=a \bar{u}+c \bar{v},\quad
u=\frac{b}{\lambda}, \quad v=\frac{d}{\lambda} . 
\ee 
Therefore, the data of $(x_{v_1e_2}^1,x_{v_1e_2}^2,y_{v_1e_2}^2)$ in (\ref{gvenew}) can be obtained by equating $g_{v_1,e_2}$ to the upper triangular matrix in (\ref{decomp}). The following steps are performed to fix the $\Su$ gauge freedom:
\begin{itemize}
    \item Compute the upper triangular matrix like (\ref{gvenew}) for $e_2$ in $v_1$ by using (\ref{decomp}). We obtain the new $\Slc$ solution for $e_2$ in both $v_1$ and $v_2$
    \be 
    \tilde{g}_{v_1e_2}=g_{v_1e_2} (h_{v_1e_1})^{-1}, \quad \tilde{\tilde{g}}_{v_2 e_2}=\tilde{g}_{v_2e_2} (h_{v_1e_1})^{-1}, 
    \ee 
    where $\tilde{g}_{v_1e_2}$ is upper triangular.
    \item The corresponding new bivector for faces of $e_2$ in $v_1$ and $v_2$ is obtained by 
    \be 
    \tilde{\tilde{X}}_{e_2 f}= h_{v_1e_1}\tilde{X}_{e_2 f}(h_{v_1e_1})^{-1}, 
    \ee
    \item Once we have $\tilde{\tilde{X}}_{e_2f}$ in $v_1$ and $v_2$, we can again use computation in Section \ref{xicomputation} to compute the corresponding $(S_{\Su})_{e_2}$ and $\xi_{e_2f}$. 
\end{itemize}

\begin{figure}[h]
	\centering
	\includegraphics[scale=0.5]{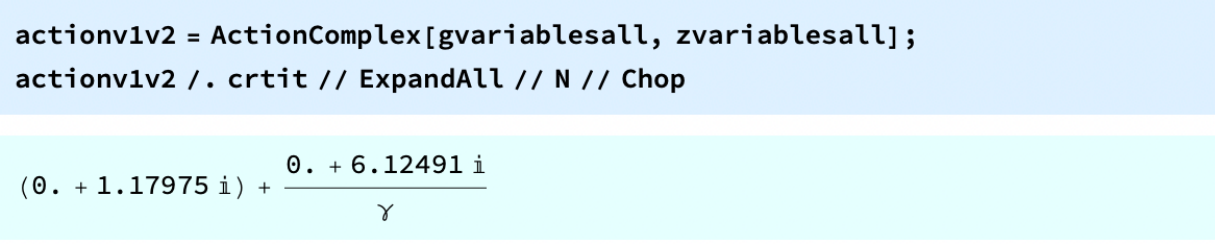}
 \caption{Computation of the action at the critical point for the complex containing two 4-simplices sharing one spacelike tetrahedron.} \label{fig:actionv1v2}
\end{figure}

\begin{figure}[h]
	\centering
	\includegraphics[scale=0.4]{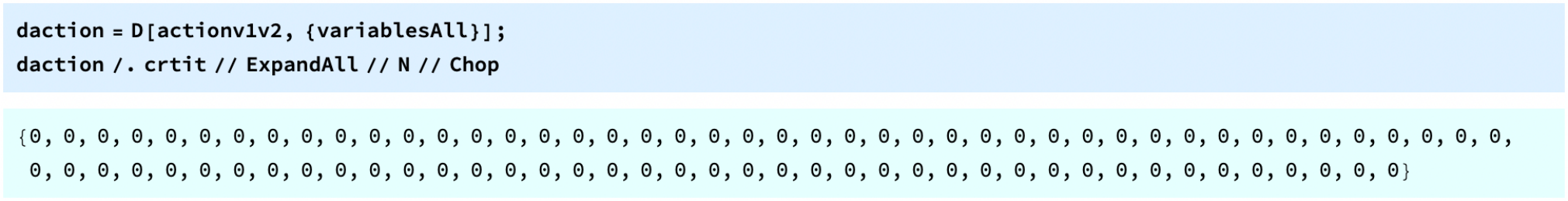}
 \caption{Computation of derivative of action at the critical point for the complex containing two 4-simplices sharing one spacelike tetrahedron. {\textbf{variableAll} does not include the spins.}} \label{fig:dactionv1v2}
\end{figure}

The above procedure construct a real critical point of the spinfoam amplitude on the complex with two 4-simplices. We fix $g_{v_1e_1}$ and $g_{v_2e_1}$ to be constant for the $\Slc$ gauge fixing. In Fig. \ref{4simplex2}, all faces of this simplicial complex are on the boundary, so the amplitude does not have the sum over $j_f$. For the timelike tetrahedra, both the timelike and spacelike boundary faces follow the parametrizations and actions given in Section \ref{SFAmplitude}. We express the action with 85 real integration variables. The action evaluated at the critical point constructed above is shown in FIG. \ref{fig:actionv1v2}. Moreover, we can verify the critical equations at the critical points, as shown in FIG. \ref{fig:dactionv1v2}.

\subsection{Internal tetrahedron is timelike}

\begin{figure}[h]
	\centering
	\includegraphics[scale=0.12]{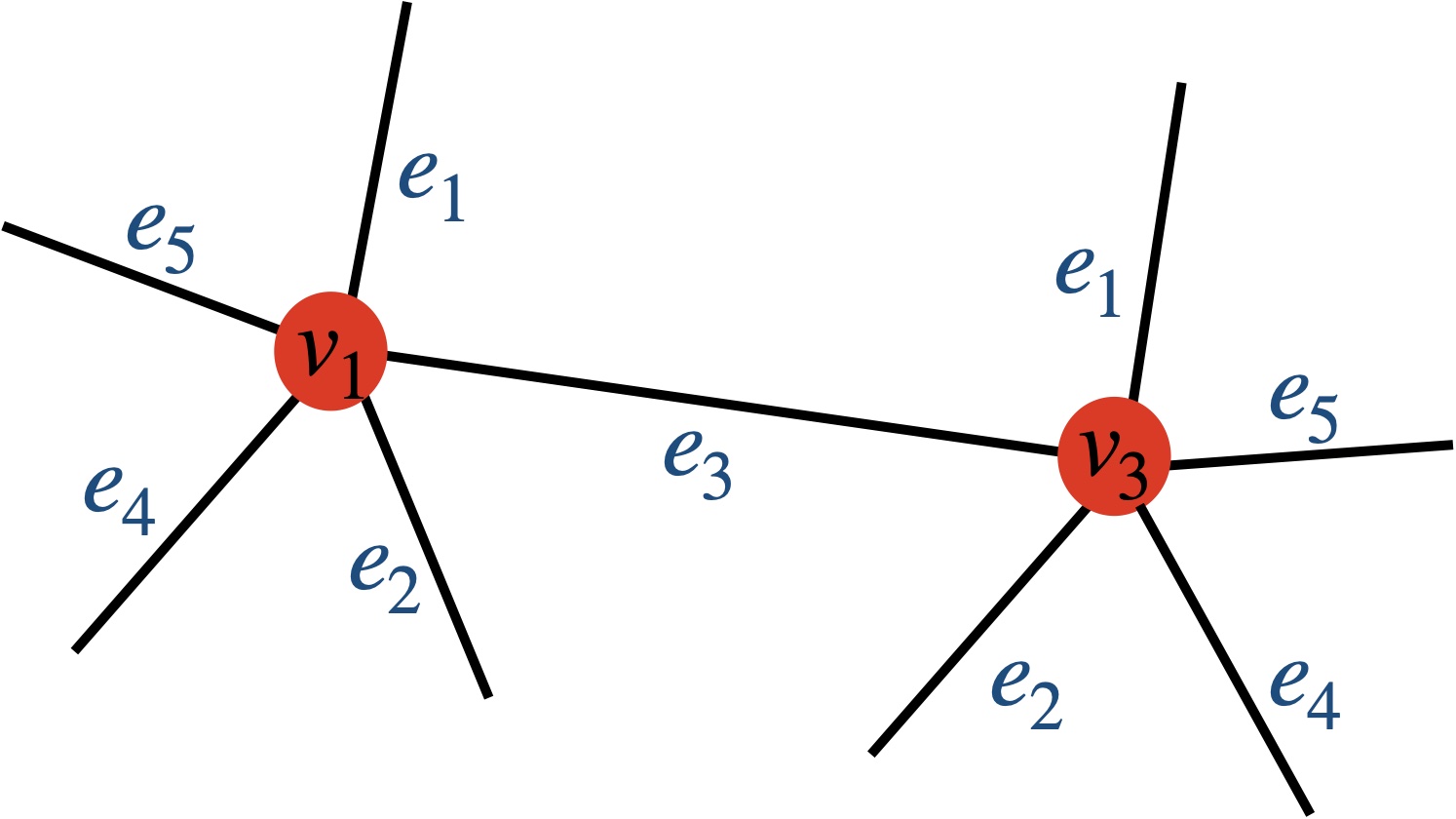}\label{4simplex2p}
 \caption{Dual diagram for two 4-simplices share one bulk timelike tetrahedron.}
\end{figure}

In this example, we will use the following coordinates for vertices the second 4-simplex, which now we denote by $v_3$.
\be 
P_1 = (0,0,0,0),\quad
P_2 = (0,0,0,1),\quad
P_{3'} = (0,1,0,1),\quad
P_4 = (0,1,1,1),\quad
P_5 = (\frac{1}{2},1,1,1).\label{coordinates3p}
\ee 
These coordinates differ from those in (\ref{vertices}) in the coordinates of $P_3$ and $P_{3'}$. The tetrahedra in $v_3$ are given by:
\be
\{v_3, e_i\}=\{\{1,2,3',4\},\{1,2,3',5\},\{1,2,4,5\},\{1,3',4,5\},\{2,3',4,5\}\}_{v_3}. \label{vertices2p}
\ee
The internal tetrahedron denoted by $e_{3}={1,2,4,5}$ (refer to Figure.\ref{4simplex2p}) is a timelike tetrahedron with 2 spacelike face and 2 timelike faces. To obtain the corresponding boundary data $\{j_{f}, \xi_{ef}, \xi^{\pm}_{ef}, l^\pm_{ef}\}$ and critical point $\{g_{v_3e},z_{v_3f}\}$ for this 4-simplex $v_3$, we repeat the procedures outlined in section \ref{v1data} starting from the coordinates in (\ref{coordinates3p}). 

\subsubsection{Match face normals}
To glue two 4-simplices, the faces of their internal tetrahedron must have matching $(j_f, \xi^{\pm}_{e_3f}, l^\pm_{e_3f})$ data, with $j_f$ already being matched. Then we exam if $(\xi^{\pm}_{e_3f}, l^\pm_{e_3f})$ in $v_1$ and $v_3$ align, and if not, we use the $(\xi^{\pm}_{e_3f}, l^\pm_{e_3f})$ in $v_1$ as the reference and adjust the corresponding $(\xi^{\pm}_{e_3f}, l^\pm_{e_3f})$ in $v_3$. The following steps are performed to align $(\xi^{\pm}_{e_3f}, l^\pm_{e_3f})$:
\begin{itemize}
    \item Firstly, we align the 3d normals $n_{e_3f}$ with each face of the shared tetrahedron $e_3$ in $v_{1}$ and $v_{3}$. We can determine an $\text{SO}(1, 2)$ matrix $O_{e_3}$ to rotate $-n_{e_3f}(v_3)$ to match $n_{e_3f}(v_1)$ by solving the linear equations $n_{e_3f}(v_1)=O_{e_3}(-n_{e_3f}(v_3))$ \footnote{We denote by $P_{e_3,ij}$ the edge vectors of $e_3$ obtained from the coordinates \eqref{coordinates3p} or equivalently from \eqref{vertices}. We have $\L_{v_3 e_3}^{-1}(P_{e_3,ij},N_{v_3e_3})=(P^{(3)}_{e_3,ij}(v_3),\hat{z})$ and $\L_{v_1 e_3}^{-1}(P_{e_3,ij},N_{v_1e_3})=(P^{(3)}_{e_3,ij}(v_1),\hat{z})$. $N_{v_1e_3}=-N_{v_3e_3}$ gives that $\L_{v_3 e_3}^{-1}\mathrm{diag}(1,1,1,-1)\L_{v_1 e_3}(P^{(3)}_{e_3,ij}(v_1),\hat{z})=(P^{(3)}_{e_3,ij}(v_3),\hat{z})$. Since $\L_{ve}\in \mathrm{SO}^+(1,3)$, it implies that $P^{(3)}_{e_3,ij}(v_1)=-O_{e_3}P^{(3)}_{e_3,ij}(v_3)$ with a certain SO(1,2) matrix $O_{e_3}$. Therefore, we obtain $n_{e_3f}(v_1)=-O_{e_3}n_{e_3f}(v_3)$, since the sets of 3d normals $n_{e_3f}(v_3)$ and $n_{e_3f}(v_1)$ are determined respectively by $P^{(3)}_{e_3,ij}(v_3)$ and $P^{(3)}_{e_3,ij}(v_1)$ by using \eqref{orthogonal}.}. In FIG. \ref{fig:SO12}, we illustrate the computation of $O_{e_3}$ using three inputs: {\bf -nabtest[[2,3]]} being 3d face normals $-n_{e_3f}(v_3)$, and {\bf nabtest[[2,3]]} being $n_{e_3f}(v_1)$, and the type of the tetrahedron $-1$. 
    \begin{figure}[h]
        \centering
        \includegraphics[scale=0.5]{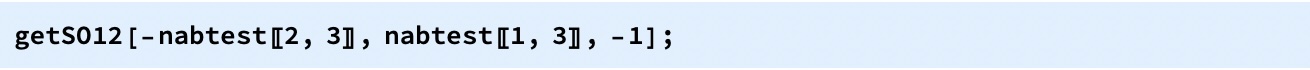}
        \caption{Computing the SO(1,2) matrix to align the 3d normals}
        \label{fig:SO12}
    \end{figure}
    
    \item The $\text{SO}(1,2)$ matrix can be embedded into a $\text{SO}^+(1, 3)$ matrix $\tilde{O}_{e_3} =  \left(\begin{array}{*{4}{c}}
    & & & 0 \\
    & & & 0\\
    \multicolumn{3}{c}
{\raisebox{\dimexpr\normalbaselineskip+.3\ht\strutbox-.3\height}[0.5pt][0.5pt]{$-O_{e_3}$}}  & 0\\
    0 & 0 & 0 & -1 \\
    \end{array}\right)$. We can then obtain the new $\text{SO}(1, 3)$ solution $\tilde{\Lambda}_{v_3 e_3}=\tilde{O}_{e_3}\Lambda_{v_3 e_3}$ \footnote{The critical equation of spinfoam action implies the parallel transport of bivectors $\L_{ve}\mathbf{X}_{ef}\L_{ve}^{-1}$ \cite{Barrett:2009mw,Han:2011re,Kaminski:2017eew,Liu:2018gfc}, which only determines $\L_{ve}$ up to a sign. But we can always fix this sign by requiring $\L_{ve}\in\mathrm{SO}^+(1,3)$, in order to lift $\L_{ve}$ to $\Slc$. }.

    \item We lift $\tilde{\Lambda}_{v_3 e_3}$ to the corresponding $\Slc$ solutions $\tilde{g}_{v_3e_3}$ by solving the equations 
    \be
    \left(\tilde{\Lambda}_{v_3e_3}^{-1}\right)^{I}_{\ J} = \frac{1}{2} \Tr\left(\tilde{g}_{v_3e_3}\bar{\sigma}^I \tilde{g}_{v_3e_3}^\dagger\bar{\sigma}^J\right),\quad \bar{\sigma}^I = (\mathbb{1}_{2\times2}, -\vec{\sigma}). \label{getsl2c1}
    \ee
    with the condition that $\det \tilde{g}_{v_3e_3}=1$. We modify $\mathring{\fg}_{v_2e_2}$ to $\mathring{\fg}_{v_2e_2}=(\tilde{g}_{v_2e_2}^T)^{-1}$.
    
    \item The difference between the previous solution $g_{v_3e_3}$ and the new solution $\tilde{g}_{v_3e_3}$ is given by
    \be 
    \delta g_{v_3e_3}=(g_{v_3e_3})^{-1}\tilde{g}_{v_3e_3}.
    \ee

    \item We obtain the new bivector $\tilde{X}_{e_3f}$ in $v_3$ by applying the following transformation:
    \be
    \tilde{X}_{e_3f}= (\delta g_{v_3 e_3})^{-1} X_{e_3 f} \delta g_{v_3e_3}.
    \ee
    \item Once we have $\tilde{X}_{e_3f}$ in $v_3$, we use the procedure in Sec.\ref{xicomputation} to determine the corresponding $(S_{\Suone})_{e_3}$ and $\left(\tilde{\xi}_{e_3f}^\pm, \tilde{l}^\pm_{e_3f}\right)$ in $v_3$. The spinor $\tilde{\xi}_{e_3f}$ in the 4-simplex $v_3$ is the same as $\xi_{e_3f}$ in the 4-simplex $v_1$. 
\end{itemize}

\subsubsection{$\Suone$ gauge fixing}

Recall the discussion about the SU(1,1) gauge fixing from (\ref{gaugesu11}) to \eqref{GFform}. For the bulk edge $e_3$, we choose a spacelike triangle, and its spinor is $\xi^+_{e_3f_2}=\left(\begin{matrix}1.017\\0.131+\imath 0.131\end{matrix}\right)$. We then find a $\Suone$ gauge transformation $U_{e_3} = \left(\begin{matrix}
    1.017 & -0.131+\imath 0.131\\ -0.131-\imath 0.131 & 1.017\end{matrix}\right)$ to transform $\xi^+_{e_3f_2}\rightarrow \left(\begin{matrix}1\\0\end{matrix}\right)$. The transformation $U_{e_3}$ acts on all $\xi^\pm_{e_3f},l^\pm_{e_3f}$ in $e_3$. Then, we choose a timelike face and rewrite the corresponding $l^+_{e_3f_4}=\left(\begin{matrix}0.62 + \imath 0.024\\0.62 - \imath 0.024\end{matrix}\right)\rightarrow \left(\begin{matrix} e^{i0.038}\\e^{-i0.038}\end{matrix}\right)$ using the scaling symmetry. Subsequently, we apply a further $\Suone$ gauge transformation $\tilde{U}_{e_3} =\left(\begin{matrix} e^{-i0.038/2} & 0\\
    0 & e^{i0.038/2}\end{matrix}\right)$. This matrix $\tilde{U}_e$ fixes $l^+_{e_3f_4}$ to $(1,1)^{\rm T}$ and again acts on all $\xi_{e_3f}^\pm$ and $l_{e_3f}^\pm$ within the same tetrahedron. Specifically, for the spacelike $f_2$, $\xi_{e_3f_2}^+$ becomes $\xi^+_{e_3f_2}= (e^{i 0.038/2},0)^{\rm T}$, and the phases can be further removed by the gauge transformation of $\xi_{ef}^\pm$. These gauge transformations allow us to fix the timelike face $l^\pm_{eh}$ in the following form: 
    \be
    l^+_{e_3f_4}= \left(\begin{matrix}
		1\\1
	\end{matrix}\right), \qquad 
    \xi^+_{e_3f_2}= \left(\begin{matrix}
		1\\
		0
	\end{matrix}\right)
	\ee 
    Then, we have the new $\xi_{e_3f_i}$ after the gauge fixing. In the end, the new $\Slc$ solutions are 
\be
\tilde{g}_{v_1 e_3} = g_{v_1 e_3}.U^{-1}_{e_3}.\tilde{U}^{-1}_{e_3}, \quad \tilde{g}_{v_3 e_3} = g_{v_3 e_3}.U^{-1}_{e_3}.\tilde{U}^{-1}_{e_3}. 
\ee 

\begin{figure}[h]
	\centering
	\includegraphics[scale=0.5]{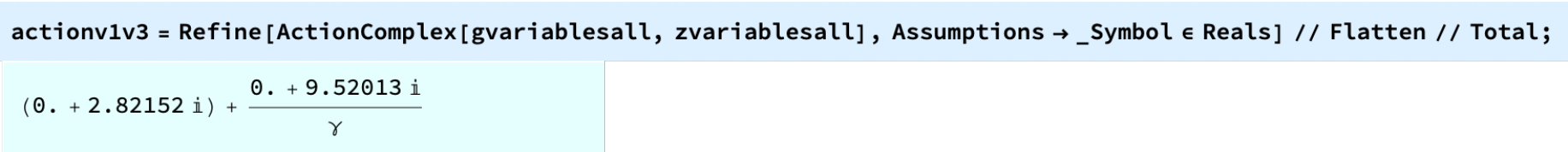}
 \caption{Computation of the action at the critical point for the simplicial complex with two 4-simplices sharing one time tetrahedron.} \label{fig:actionv1v3}
\end{figure}

\begin{figure}[ht]
	\centering
	\includegraphics[scale=0.5]{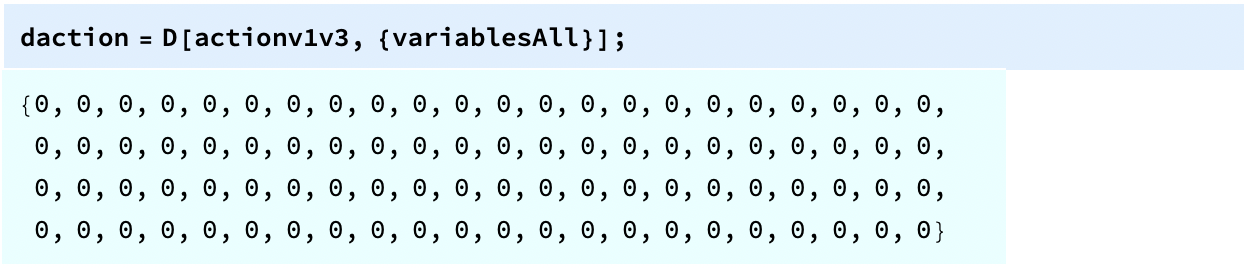}
 \caption{Computation of the derivative of the action at the critical point for the simplicial complex with two 4-simplices sharing one timelike tetrahedron.} \label{fig:dactionv1v3}
\end{figure}

The $\Slc$ gauge fixing is the same as in a single 4-simplex: We fix $\fg_{v_1e_1}=\mathring{\fg}_{v_1e_1}$ and $\fg_{v_3e_1}=\mathring{\fg}_{v_3e_1}$ in $v_1$ and $v_3$ respectively. The parametrization and action have been provided in Section \ref{SFAmplitude}. Using the coordinates given by (\ref{vertices}) and (\ref{vertices2p}), we express the action as a function depending on 88 real variables. The evaluation of the action at the critical point is shown in FIG. \ref{fig:actionv1v3}. Moreover, we can verify the critical equations at the critical points, as illustrated in FIG. \ref{fig:dactionv1v3}.

\subsection{Compare to Regge action} \label{Sec:Regge2}

So far, we have considered the real critical point $\mathring{x}=\{\mathring{\fg}_{ve}, \mathring{z}_{vf}\}$ with the orientation $(s_{v_1},s_{v_2})=(+,+)$ on the complex. With the given boundary data $\mathring{r}=\{j_f, \xi_{ef},\xi^\pm_{ef},l^+_{ef}\}$, there are exactly 4 real critical points corresponding to the orientations $(s_{v_1},s_{v_2})=(+,+),(+,-),(-,+),(-,-)$. The $\Slc$ solutions for the other orientations can be computed using (\ref{gveminus}) based on the $\Slc$ solution obtained above for the orientation $(s_{v_1},s_{v_2})=(+,+)$. Consequently, one can obtain the solutions of spinors $\mathring{z}^{s_{v_1}s_{v_2}}_{vf}$ for different orientations. Then we can evaluate the spinfoam action $S^{(++)},S^{(+-)},S^{(-+)},S^{(--)}$ at real critical points for different orientations. 

The spinfoam action involves an overall phase $\phi$ related to the boundary data, which can be determined by: 
\be 
\phi =\frac{1}{2}\left(S^{(++)}+S^{(--)}\right). 
\ee 
Excluding this arbitrary phase, we can compare the spinfoam action to the Regge action:
\be
S^{s_{v_1}s_{v_2}}_{\rm Regge} = \sum_{k=1}^2 s_{v_k}\left(\sum_{i=1}^5 \left|A_{f}\right|\tilde{\theta}_{f}(v_k)\right). \label{eq:regge2}
\ee 
Here, $s_{v_k}=\pm1$ depends on the orientation of the $k$-th 4-simplex on the simplicial complex, $\left|A_{f}\right|$ represents the absolute value of the area of face $f$, and $\tilde{\theta}_{f}(v_k)$ represents the dihedral angle hinged by $f$ in the $k$-th 4-simplex. For spacelike faces, $\tilde{\theta}_{f}(v_k) = \theta_{e_ie_j}$, and for timelike faces, 
$\tilde{\theta}_{f}(v_k) =\pi- \theta_{e_ie_j}$. Here, $\theta_{e_ie_j}$ with $f=e_i\cap e_j$ in $v_k$ is defined in Sec \ref{dihedralAnglesSec}. According to \cite{Kaminski:2017eew,Liu:2018gfc,Han:2021rjo}, the difference between the spinfoam action (excludes the phase) and Regge action has the relation that 
\be 
\delta S^{s_{v_1}s_{v_2}}& =& S^{s_{v_1}s_{v_2}} - \imath S^{s_{v_1}s_{v_2}}_{\rm Regge} \nonumber\\
&=& \sum_{f_1\in t\text{-}s} \left(\pm\imath  \frac{\pi}{2}\right)j_{f_1}+\sum_{f_2\in t\text{-}s} \left(\pm\imath {\pi}\text{ or } 0\right)j_{f_2} + \sum_{f\in t\text{-}t} \left(\imath \pi\text{ or } 0\right) j_f. \label{dSdregge2}
\ee 
$f_1\in t$-$s$ denotes the spacelike triangle involving only one timelike tetrahedron. $f_2\in t$-$s$ denotes the spacelike face shared by two timelike tetrahrdra. In our examples, the spacelike internal tetrahedron has a pair of triangles of the type $f_2$. $f\in t\text{-}s$ denotes the timelike face. The numerical results of $\delta S^{s_{v_1}s_{v_2}}$ in FIG. \ref{fig:dsdRegge2} and FIG. \ref{fig:dsdRegge3} confirm (\ref{dSdregge2}).

\begin{figure}[h]
    \centering\caption{The numerical result of $\delta S^{s_{v_1}s_{v_2}}$ for the case that $v_1$ and $v_2$ share a spacelike tetrahedron $e_2$. All the spins $j_f\in {\mathbb{N}_0}/{2}$. The spacelike triangles carrying $\text{j1$\$$1$\$$2},\text{j1$\$$1$\$$4},\text{j1$\$$3$\$$1},\text{j1$\$$5$\$$1},\text{j2$\$$1$\$$3},\text{j2$\$$3$\$$2},\text{j2$\$$3$\$$5}$, and $\text{j2$\$$4$\$$3}$ involve only one timelike tetrahedron and thus belong to type $ f_1$ in (\ref{dSdregge2}). The spacelike triangles carrying $\text{j1$\$$4$\$$2}$ and  $\text{j2$\$$5$\$$2}$ involve two timelike tetrahedra and thus belong to type $ f_2$ in (\ref{dSdregge2}). The timelike triangles carry $\text{j1$\$$3$\$$4},\text{j1$\$$4$\$$5},\text{j2$\$$1$\$$5}$, and $\text{j2$\$$5$\$$4}$.}
    \includegraphics[scale=0.35]{figures/dsdsregge2.pdf}
    \label{fig:dsdRegge2}
\end{figure}

\begin{figure}[h]
    \centering
    \caption{The numerical result of $\delta S^{s_{v_1}s_{v_2}}$ for the case that $v_1$ and $v_3$ share a timelike tetrahedron $e_3$. All the spins $j_f\in {\mathbb{N}}/{2}$. The spacelike triangles carrying $\text{j1$\$$1$\$$4},\text{j1$\$$2$\$$5},\text{j1$\$$4$\$$2},\text{j1$\$$5$\$$1},\text{j2$\$$1$\$$5},\text{j2$\$$2$\$$4},\text{j2$\$$4$\$$1}$, and $\text{j2$\$$5$\$$2}$ involve only one timelike tetrahedron and thus belong to type $ f_1$ in (\ref{dSdregge2}). The spacelike triangles carrying $\text{j1$\$$2$\$$3}$ and  $\text{j2$\$$1$\$$3}$ involve two timelike tetrahedra and thus belong to type $ f_2$ in (\ref{dSdregge2}). The only timelike triangle carries $\text{j1$\$$5$\$$3}$. }
    \includegraphics[scale=0.4]{figures/dsdsregge3.pdf}
    \label{fig:dsdRegge3}
\end{figure}

\section{Complex critical points}

\subsection{Real and Complex Critical Points} \label{CCP}
 
By the stationary phase approximation, the dominant contributions to the integrals (\ref{SFamplitude}) come from the critical points satisfying the critical equations. The critical points inside the integration domain, denoted by $\{\mathring{j}_h,\mathring{X}\}$, satisfy the following critical equations from the spinfoam action $S$:
\be
\re(S_{\rm SF})&=&\partial_{X}S_{\rm SF}=0,\label{eom1}\\
\partial_{j_f}S_{\rm SF}&=&4\pi i k_f, \qquad k_f\in\Z.\label{eom2}
\ee 
We view the integration domain as a real manifold, and call $\{\mathring{j}_h,\mathring{X}\}$ the \textit{real critical point}. The critical points constructed by the above algorithm are all real. As we have seen in the above, the real critical point closely relates to the nondegenerate Regge geometry. Eq.\eqref{eom2} further imposes a curvature constraint to the geometry \cite{Bonzom:2009hw,Han:2013hna,Hellmann:2012kz,Perini:2012nd,Engle:2020ffj}. Consequently, the existence of a real critical point depends on the boundary conditions and may not hold for generic conditions \cite{Han:2023cen,Han:2021kll}. To analyze the asymptotics of the amplitude in absence of real critical point, we apply the stationary phase approximation for complex action with parameters \cite{10.1007/BFb0074195,Hormander} and compute the \textit{complex critical points}. The generic curved spacetime geometry always relates to the complex critical point rather than the real critical point \cite{Han:2021kll}.

\begin{figure}[h]
    \centering
    \includegraphics[scale=0.15]{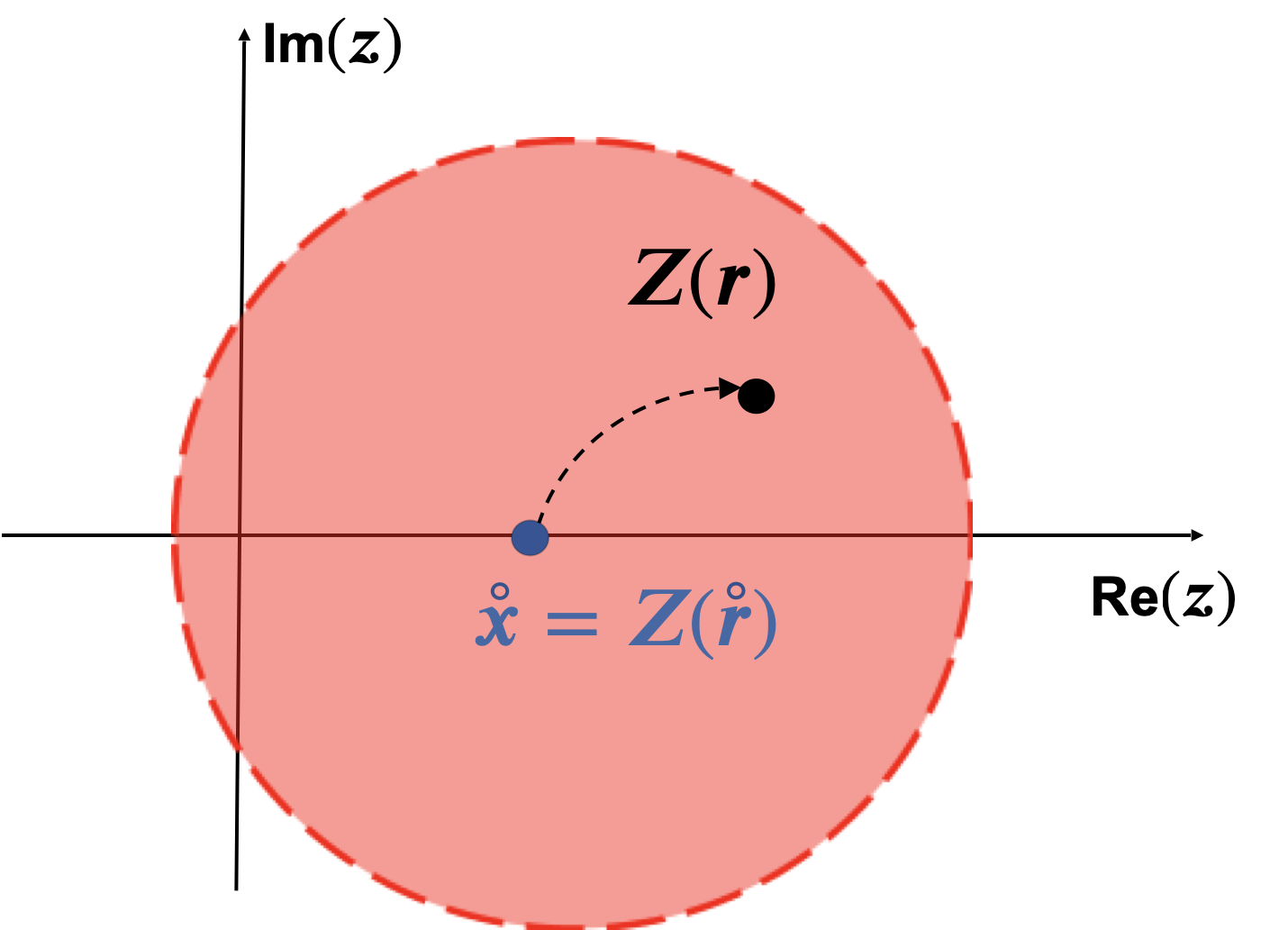}
    \caption[Caption for LOF0]{The real and complex critical points $\mathring{x}$ and $Z(r)$. $\cs(r,z)$ is analytic extended from the real axis to the complex neighborhood illustrated by the red disk.}
    \label{Figure0}
\end{figure}

We review briefly the scheme of analysis, we consider the large-$\lambda$ integral 
\be 
\int_K e^{\lambda S(r,x)}\mathrm{d}^N x,
\ee
where $r$ denotes the external parameters, and $S(r,x)$ is an analytic function of $r\in U\subset \R^k$ and $x\in K\subset \R^N$. Here, $U\times K$ forms a neighborhood of $(\mathring{r},\mathring{x})$, where $\mathring{x}$ is a real critical point of $S(\mathring{r},x)$. We denote the analytic extension of $S(r,x)$ to a complex neighborhood of $\mathring{x}$ as $\mathcal{S}(r,z)$, with $z=x+iy \in \mathbb{C}^{N}$. For generic $r\neq\mathring{r}$, the complex critical equation 
\be 
\partial_{z} \mathcal{S}=0,\label{criticaleqn111}
\ee
gives the solution $z=Z(r)$, which generically moves away from the real plane $\R^N$. Therefore, we call $Z(r)$ the {complex critical point} (see FIG.\ref{Figure0}). 
The large-$\lambda$ asymptotic expansion for the integral can be established with the complex critical point:
\be
\int_K e^{\lambda S(r,x)}  \mathrm{d}^N x = \left(\frac{2\pi}{\lambda}\right)^{\frac{N}{2}} \frac{e^{\lambda \mathcal{S}(r,Z(r))}}{\sqrt{\det(-\mathcal{S}_{z,z}(r,Z(r)))}} \lt[1+O(1/\l)\rt].\label{asymptotics0}
\ee
Here, $\mathcal{S}(r,Z(r))$ and $\cs_{z,z}(r,Z(r))$ are the action and Hessian evaluated at the complex critical point. Furthermore, the real part of $\mathcal{S}$ satisfies the condition:
\be
\operatorname{Re}(\mathcal{S}) \leq-C|\operatorname{Im}(Z)|^{2}.\label{negativeReS}
\ee 
where $C$ is a positive constant. We refer to \cite{10.1007/BFb0074195,Hormander} for a detailed proof of this inequality. This condition implies that in \eqref{asymptotics0}, the oscillatory phase can only occur at the real critical point, where $\operatorname{Im}(Z)=0$ and $r=\mathring{r}$. When $r$ deviates from $\mathring{r}$, causing $\operatorname{Im}(Z)$ finite and $\operatorname{Re}(\mathcal{S})$ to become negative, the result in \eqref{asymptotics0} is exponentially suppressed as $\lambda$ grows large. Nevertheless, we can arrive at a regime where the asymptotic behavior described in \eqref{asymptotics0} is not suppressed at the complex critical point. In fact, for any sufficiently large $\lambda$, we can always find a value of $r$ close to but not equal to $\mathring{r}$. In this region, both $\operatorname{Im}(Z)$ and $\mathrm{Re}(\mathcal{S})$ can be made small enough, ensuring that $e^{\lambda \cs}$ in \eqref{asymptotics0} is not significantly suppressed at the complex critical point.

\subsection{Real critical point for $\Delta_3$ triangulations}

\begin{figure}[ht]
	\centering
	\includegraphics[scale=0.3]{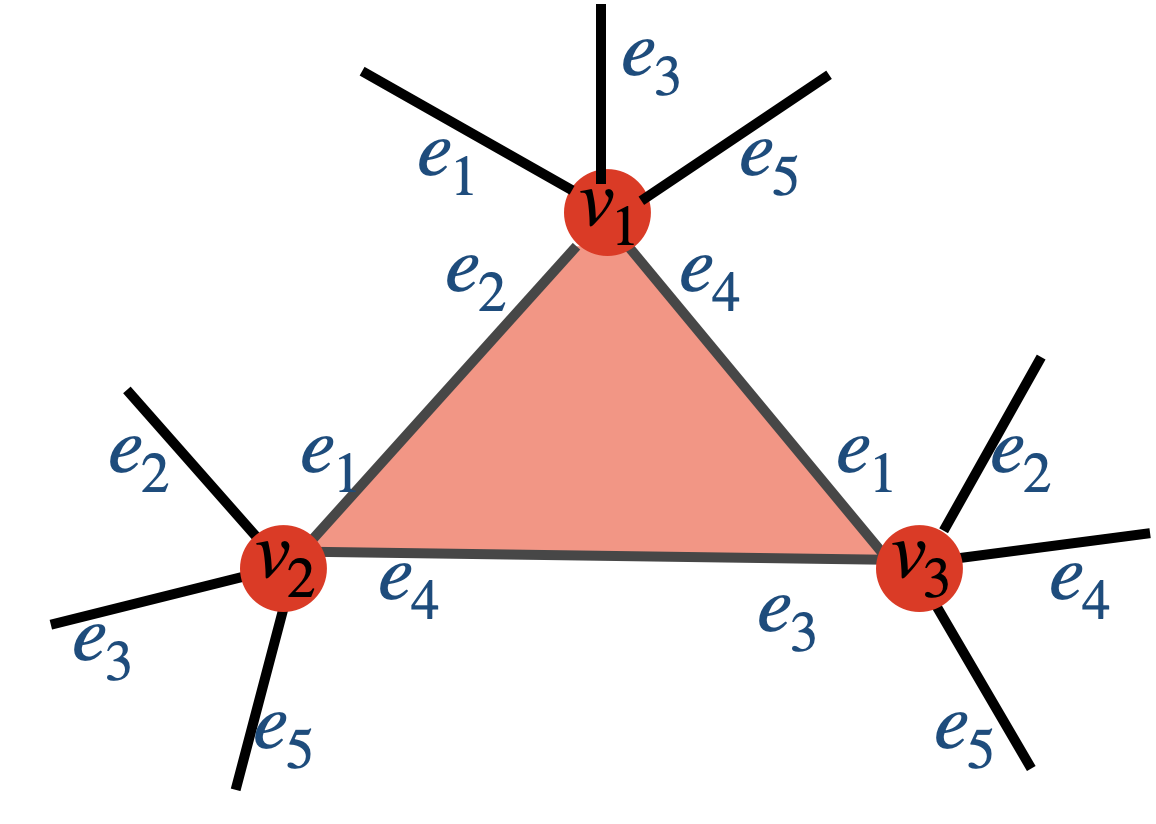}
 \caption{Dual diagram for $\Delta_3$ triangulation, which shares one spacelike bulk face} \label{4simplex3}
\end{figure}

\begin{figure}[ht]
	\centering
	\includegraphics[scale=0.5]{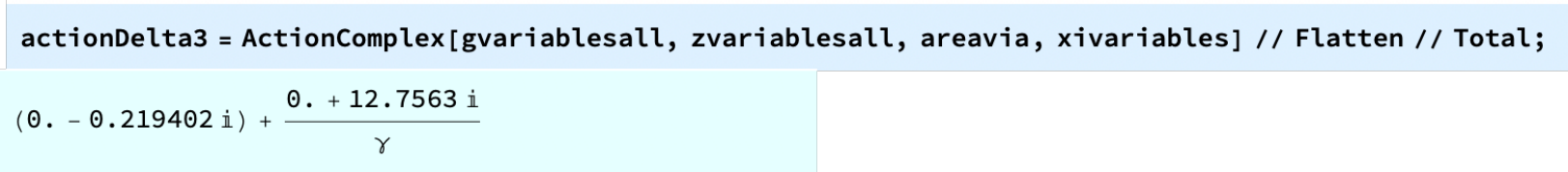}
 \caption{Evaluate the action at the real critical points for the $\Delta_3$ triangulation.} \label{fig:actionDelta3}
\end{figure}

\begin{figure}[h]
	\centering
	\includegraphics[scale=0.5]{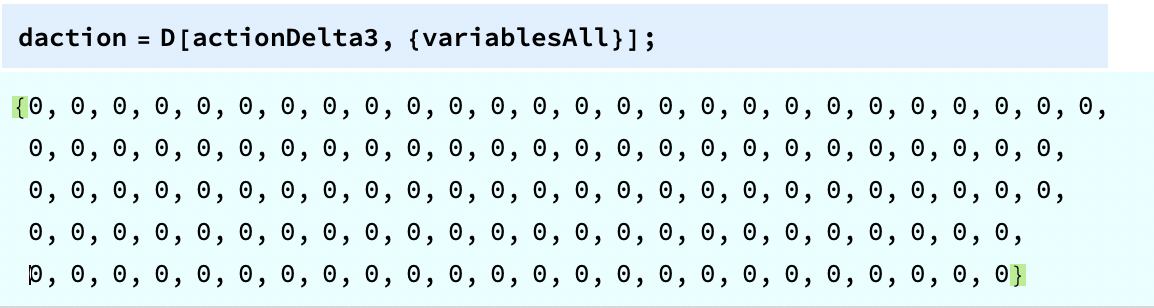}
 \caption{The derivative of the action at the critical point for the $\Delta_3$ triangulation.} \label{fig:dactionDelta3}
\end{figure}

The $\Delta_3$ triangulation consists of three 4-simplices sharing a common triangle. The tetrahedra in this simplicial complex are given by
\be 
\begin{gathered}
    \{v_1, e_i\}=\{\{1,2,3,4\},\{1,2,3,5\},\{1,2,4,5\},\{1,3,4,5\},\{2,3,4,5\}\}_{v_1},\\
\{v_2, e_i\}=\{\{1,2,3,5\},\{1,2,3,6\},\{1,2,5,6\},\{1,3,5,6\},\{2,3,5,6\}\}_{v_2},\\
\{v_3, e_i\}=\{\{1,3,4,5\},\{1,3,4,6\},\{1,3,5,6\},\{1,4,5,6\},\{3,4,5,6\}\}_{v_3}.
\end{gathered}
\ee 
The internal face is $h=(1,3,5)$, see FIG. \ref{4simplex3}. 
The flat Regge geometry on $\Delta_3$ is determined by the following coordinates of the vertices in $(\R^4,\eta)$
\be
& P_{1}=(0,0,0,0),\quad P_{2}=\left(0,-2 \sqrt{10} / 3^{3 / 4},-\sqrt{5} / 3^{3 / 4},-\sqrt{5} / 3^{1 / 4}\right),\nonumber\\ & P_{3}=\left(0,0,0,-2 \sqrt{5} / 3^{1 / 4}\right), \quad P_{4}=\left(-3^{-1 / 4} 10^{-1 / 2},-\sqrt{5 / 2} / 3^{3 / 4},-\sqrt{5} / 3^{3 / 4},-\sqrt{5} / 3^{1 / 4}\right),\nonumber\\ & P_{5}=\left(0,0,-3^{1 / 4} \sqrt{5},-3^{1 / 4} \sqrt{5}\right), \quad P_6=\left(0.90, 2.74, -0.98, -1.70 \right). \label{P6}
\ee
Then the procedure outlined in sections \ref{v1data} and \ref{SimplicialComp} constructs the corresponding boundary data $(j_{b}, \xi_{ef})$ and the real critical point $(\mathring{j}_h,\mathring{\fg}_{v_e},\mathring{z}_{vf})$ for the spinfoam amplitude on $\Delta_3$. The action $S$ is expressed with 124 real variables and evaluated at the real critical point, as shown in FIG. \ref{fig:actionDelta3}. Moreover, we verify the critical equations at the real critical points, as illustrated in FIG. \ref{fig:dactionDelta3}. Here the critical equations include the derivative of the action with respect to $j_h$.

\subsection{Parity transform and Compare to Regge action}

Given the boundary data $\mathring{x}=\{\xi_{eb},j_b\}$ on the $\Delta_3$ triangulation, there are exactly two real critical points $\mathring{x}$ and $\mathring{x}'$, where $\mathring{x}$ and $\mathring{x}'$ correspond to the same flat geometry with the orientations of 4-simplices $(s_{v_1},s_{v_2},s_{v_3})=(+,+,+)$ or $(-,-,-)$. Other 6 discontinuous orientations $(+,+,-),(+,-,+),(+,-,-),(-,+,+),(-,+,-),(-,-,+)$ do not result in any real critical point, because they all violates the flatness constraint $\g {\delta}^{s}_h=\g\sum_vs_v\theta_h(v)=0$. The magnitude of $|{\delta}^{s}_h|$ is not negligible for the discontinuous orientation, so the contribution to $A(\Delta_3)$ is suppressed even when considering the complex critical point. The above calculation is based on the orientation $(s_{v_1},s_{v_2},s_{v_3})=(-,-,-)$, and the real critical points $\mathring{x}'=\{\mathring{j}_h,\mathring{\fg}^{(+++)}_{ve},\mathring{z}^{(+++)}_{ve}\}$ for another orientation can also be computed following the procedure in Sec. \ref{Sec:Regge2}. Then, the spinfoam action $S^{(+++)}$ for the orientation $(+,+,+)$ is evaluated at the real critical point $\mathring{x}'$. The overall phase and the phase difference are given by
\be 
\phi_{\Delta_3} = \frac{1}{2}\left(S^{(+++)}+S^{(---)}\right),\qquad \Delta S_{\Delta_3}=S^{(+++)}-S^{(---)}.
\ee 
We have $S^{(\pm\pm\pm)}-\phi_{\Delta_3}=\pm\Delta S_{\Delta_3}/2$ and the asymptotics of the amplitude (when the boundary date corresponds to a flat geometry)
\be
e^{\imath \phi_{\Delta_3}}\lt(\cn_+ e^{\imath \Delta S_{\Delta_3}/2}+\cn_- e^{-\imath \Delta S_{\Delta_3}/2}\rt) \lt[1+O(1/\l)\rt]. 
\ee
We can compare the spinfoam action to the Regge action
\be 
\left(S^{(\pm\pm\pm)}-\phi_{\Delta_3}\right)=\pm \imath S_{\rm Regge},\label{eq:dsdsregge4}
\ee 
The Regge action is given by $S_{\rm Regge}=A_h\delta _h+\sum_bA_b\theta_b$, where ${\delta}_h=\sum_v\theta_h(v)$ and $\theta_b=\sum_v\theta_b(v)$ for internal triangle $h$ and boundary triangle $b$. The numerical result confirms (\ref{eq:dsdsregge4}).

\subsection{Complex critical point for $\Delta_3$ triangulations}

\begin{figure}[h]
	\centering
 \caption{Computation of the complex critical points for the $\Delta_3$ triangulation.} 
	\includegraphics[scale=0.5]{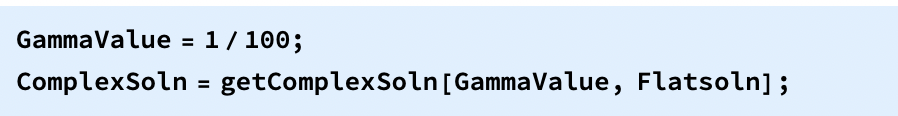}
\label{fig:complexsoln}
\end{figure}

In the case where we are given the curved Regge geometry with the coordinates $P_1$ to $P_6$ as described in (\ref{P6}), but for $v_2= (1,2,3,5,6')$ with the coordinates $P_{6'}=(0.904, 2.746, -0.981, -1.699)$, we follow the procedure outlined above to construct the boundary data $(j_b,\xi_{bf})$. It can be checked that the critical equation $\partial_{j_h} S\neq 0$ now. There is actually no real solution for the full set of critical equations. We need to solve for the complex critical points to satisfy the critical equations. According to Sec.
\ref{CCP}, the complex critical points $\textbf{ComplexSoln}$ are in the neighborhood of real critical points, we can search for the complex critical points starting from the real critical points $\mathring{x}$ by using the default function \textbf{FindRoot} in Mathematica. Moreover, the $\Slc$ gauge fixing in the curved geometry should adapt the same data $(\mathring{\fg}_{v_1e_1},\mathring{\fg}_{v_2e_2},\mathring{\fg}_{v_3e_2})$ with the flat geometry. In our code, the function \textbf{getComplexSoln} can compute the complex critical points with the input of real critical points \textbf{Flatsoln}. In FIG. \ref{fig:complexsoln}, we illustrate the computation of the complex critical points at $\gamma=0.01$.

When we have a relatively large deformation of boundary data such that the complex critical is relatively far from the real critical point, using \textbf{FindRoot} may fail to find the complex critical point accurately. Then We may need to split the large deformation into multiple steps of small deformations and use the real critical point as the starting point of \textbf{FindRoot} in the 1st step deformation. We use the complex critical point resulting from the $(n-1)$-th step deformation to be the starting point of \textbf{FindRoot} in the $n$-th step deformation. This iteration gives more accurately the complex critical point corresponding to a relatively large deformation of boundary data.

\begin{figure}[h]
\centering
\caption{Computation of the action at the complex critical point $\textbf{ComplexSoln}$ for the $\Delta_3$ triangulation.}	
\includegraphics[scale=0.5]{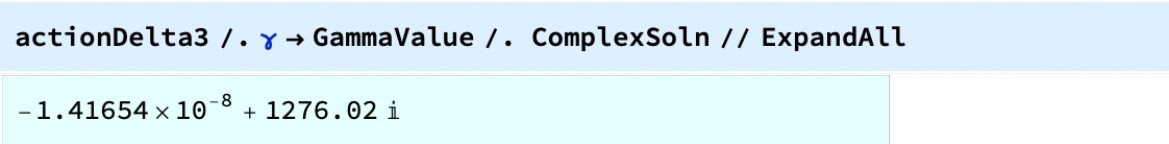}
\label{fig:actionDelta3Complex}
\end{figure}

\begin{figure}[h]
	\centering
 \caption{Computation of the derivative of the action at the complex critical point for the $\Delta_3$ triangulation.}
\includegraphics[scale=0.5]{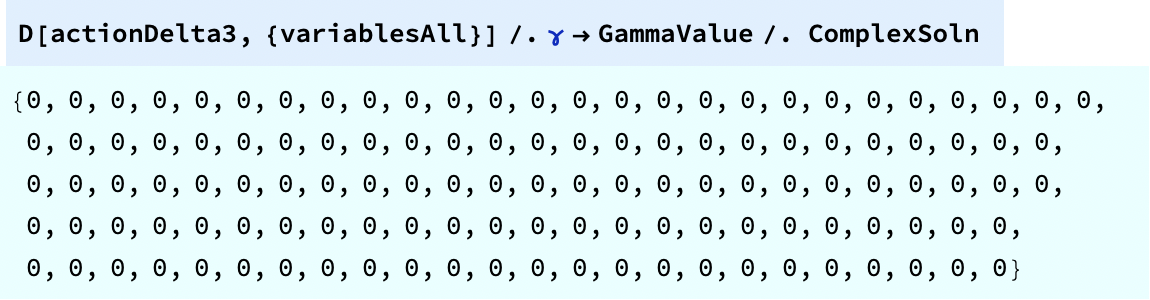}
\label{fig:dactionDelta3Complex}
\end{figure}

Next, we evaluate the action at the complex critical points $\textbf{ComplexSoln}$ when $\gamma=0.01$, as shown in FIG. \ref{fig:actionDelta3Complex}. Particularly, the solution of the internal spin is $j_h=499.999 + 0.001417\imath$. The real part of the action is negative, which also confirms (\ref{negativeReS}). Finally, we verify that the critical equations are now satisfied, as shown in FIG. \ref{fig:dactionDelta3Complex}.

\section{Conclusion and outlook} \label{Conclusion and Discussion} 

In this work, we have presented a comprehensive algorithm and corresponding Mathematica program for the construction of boundary data and computation of real and complex critical points in spinfoams. We first use a single 4-simplex as an example to illustrate the procedure to compute the boundary data $(j_f, \xi_{ef},\xi_{ef}^\pm,l^+_{ef})$ and the corresponding real critical point $(g_{ve},z_{vf})$ for a given geometry. Then, we generalize the algorithm to the simplicial complex and use two 4-simplices sharing a common internal tetrahedron as an example to illustrate how to obtain the boundary data and the real critical points on the simplicial complex. To introduce the concept of complex critical points in spinfoams, we use the $\Delta_3$ triangulation as an example to demonstrate the algorithm for computing complex critical points when the boundary data admits a curved geometry where real critical points do not exist. Additionally, we explore the parity transform in spinfoams and compare our numerical results of the spinfoam action at the real critical points with the Regge action. 

Our work provides a general procedure to numerically construct the critical points of the spinfoam amplitude on the simplicial complex. It contributes to advancing computational techniques in the study of spinfoam amplitudes and lays a foundation for continued progress and innovation in covariant Loop Quantum Gravity. Some investigations have been carried out by applying the algorithm to compute the spinfoam amplitude in the physical scenarios such as cosmology and black holes. This numerical method turns out to be powerful enough to identify the complex critical points of spinfoam amplitude on relatively large triangulations that relate to cosmology and black-hole-to-white-hole transition \cite{Han:2024ydv, Han:2024rqb}. The results may initiate a new approach to embedding cosmology and black hole models within the full theory of covariant Loop Quantum Gravity and hopefully relating to existing results from coherent path integral formulation of Loop Quantum Gravity \cite{Han:2019vpw,Han:2021cwb}, Loop Quantum Cosmology \cite{Ashtekar:2006wn} and nonsingular black holes \cite{Han:2022rsx,Ashtekar:2023cod,Giesel:2023hys,Giesel:2023reg}.

The Mathematica program demonstrated in this work focuses on the critical points and their contributions to the leading order of $1/j$ expansion. The program will be extended to the computation of next-to-leading order term in the expansion, by generalizing the earlier work \cite{Han:2020fil}. Some generalization has been carried out in studying the next-to-leading order term of the amplitude on the 1-5 Pachner move complex \cite{toappear15}.

\section*{Acknowledgments}
MH receives support from the National Science Foundation through grants PHY-2207763, the College of Science Research Fellowship at Florida Atlantic University. MH, HL and DQ are supported by research grants provided by the Blaumann Foundation. The Government of Canada supports research at Perimeter Institute through Industry Canada and by the Province of Ontario through the Ministry of Economic Development and Innovation. This work benefits from the visitor's supports from Beijing Normal University, FAU Erlangen-N\"urnberg, the University of Western Ontario, and Perimeter Institute for Theoretical Physics. 

\bibliographystyle{jhep}
\bibliography{ref}

\providecommand{\href}[2]{#2}\begingroup\raggedright\begin{thebibliography}{10}

\bibitem{Dona:2022yyn}
P.~Dona, M.~Han, and H.~Liu, {\it {Spinfoams and high performance computing}},
  \href{http://arxiv.org/abs/2212.14396}{{\tt arXiv:2212.14396}}.

\bibitem{Han:2021kll}
M.~Han, Z.~Huang, H.~Liu, and D.~Qu, {\it {Complex critical points and curved
  geometries in four-dimensional Lorentzian spinfoam quantum gravity}},  {\em
  Phys. Rev. D} {\bf 106} (2022), no.~4 044005,
  [\href{http://arxiv.org/abs/2110.10670}{{\tt arXiv:2110.10670}}].

\bibitem{Han:2023cen}
M.~Han, H.~Liu, and D.~Qu, {\it {Complex critical points in Lorentzian spinfoam
  quantum gravity: Four-simplex amplitude and effective dynamics on a
  double-\ensuremath{\Delta}3 complex}},  {\em Phys. Rev. D} {\bf 108} (2023),
  no.~2 026010, [\href{http://arxiv.org/abs/2301.02930}{{\tt
  arXiv:2301.02930}}].

\bibitem{Han:2024ydv}
M.~Han, H.~Liu, D.~Qu, F.~Vidotto, and C.~Zhang, {\it {Cosmological Dynamics
  from Covariant Loop Quantum Gravity with Scalar Matter}},
  \href{http://arxiv.org/abs/2402.07984}{{\tt arXiv:2402.07984}}.

\bibitem{Han:2024rqb}
M.~Han, D.~Qu, and C.~Zhang, {\it {Spin foam amplitude of the black-to-white
  hole transition}},  \href{http://arxiv.org/abs/2404.02796}{{\tt
  arXiv:2404.02796}}.

\bibitem{Han:2020fil}
M.~Han, Z.~Huang, H.~Liu, and D.~Qu, {\it {Numerical computations of
  next-to-leading order corrections in spinfoam large-$j$ asymptotics}},  {\em
  Phys. Rev. D} {\bf 102} (2020), no.~12 124010,
  [\href{http://arxiv.org/abs/2007.01998}{{\tt arXiv:2007.01998}}].

\bibitem{toappear15}
M.~Han, H.~Li, H.~Liu, and D.~Qu, {\it {Landscape of 4D spinfoam quantum
  geomoetry: Results from next-to-leading order spinfoam large-$j$ asymptotics
  of 1-5 Pachner move}},  \href{http://arxiv.org/abs/to appear}{{\tt to
  appear}}.

\bibitem{Gozzini:2021kbt}
F.~Gozzini, {\it {A high-performance code for EPRL spin foam amplitudes}},
  {\em Class. Quant. Grav.} {\bf 38} (2021), no.~22 225010,
  [\href{http://arxiv.org/abs/2107.13952}{{\tt arXiv:2107.13952}}].

\bibitem{Frisoni:2022urv}
P.~Frisoni, F.~Gozzini, and F.~Vidotto, {\it {Markov Chain Monte Carlo methods
  for graph refinement in Spinfoam Cosmology}},
  \href{http://arxiv.org/abs/2207.02881}{{\tt arXiv:2207.02881}}.

\bibitem{Dona:2022dxs}
P.~Dona and P.~Frisoni, {\it {How-to Compute EPRL Spin Foam Amplitudes}},  {\em
  Universe} {\bf 8} (2022), no.~4 208,
  [\href{http://arxiv.org/abs/2202.04360}{{\tt arXiv:2202.04360}}].

\bibitem{Asante:2020qpa}
S.~K. Asante, B.~Dittrich, and H.~M. Haggard, {\it {Effective Spin Foam Models
  for Four-Dimensional Quantum Gravity}},  {\em Phys. Rev. Lett.} {\bf 125}
  (2020), no.~23 231301, [\href{http://arxiv.org/abs/2004.07013}{{\tt
  arXiv:2004.07013}}].

\bibitem{Asante:2021zzh}
S.~K. Asante, B.~Dittrich, and J.~Padua-Arguelles, {\it {Effective spin foam
  models for Lorentzian quantum gravity}},  {\em Class. Quant. Grav.} {\bf 38}
  (2021), no.~19 195002, [\href{http://arxiv.org/abs/2104.00485}{{\tt
  arXiv:2104.00485}}].

\bibitem{Asante:2022lnp}
S.~K. Asante, J.~D. Sim\~ao, and S.~Steinhaus, {\it {Spin-foams as
  semi-classical vertices: gluing constraints and a hybrid algorithm}},
  \href{http://arxiv.org/abs/2206.13540}{{\tt arXiv:2206.13540}}.

\bibitem{Asante:2022dnj}
S.~K. Asante, B.~Dittrich, and S.~Steinhaus, {\it {Spin foams, Refinement limit
  and Renormalization}},  \href{http://arxiv.org/abs/2211.09578}{{\tt
  arXiv:2211.09578}}.

\bibitem{Bahr:2016hwc}
B.~Bahr and S.~Steinhaus, {\it {Numerical evidence for a phase transition in 4d
  spin foam quantum gravity}},  {\em Phys. Rev. Lett.} {\bf 117} (2016), no.~14
  141302, [\href{http://arxiv.org/abs/1605.07649}{{\tt arXiv:1605.07649}}].

\bibitem{Barrett:2009mw}
J.~W. Barrett, R.~J. Dowdall, W.~J. Fairbairn, F.~Hellmann, and R.~Pereira,
  {\it {Lorentzian spin foam amplitudes: Graphical calculus and asymptotics}},
  {\em Class. Quant. Grav.} {\bf 27} (2010) 165009,
  [\href{http://arxiv.org/abs/0907.2440}{{\tt arXiv:0907.2440}}].

\bibitem{Han:2011re}
M.~Han and M.~Zhang, {\it {Asymptotics of Spinfoam Amplitude on Simplicial
  Manifold: Lorentzian Theory}},  {\em Class. Quant. Grav.} {\bf 30} (2013)
  165012, [\href{http://arxiv.org/abs/1109.0499}{{\tt arXiv:1109.0499}}].

\bibitem{numericalResult}
H.~Liu and D.~Qu.
  \url{https://github.com/dqu2017/Real-and-Complex-Critical-Points}, 2024.

\bibitem{Engle:2007wy}
J.~Engle, E.~Livine, R.~Pereira, and C.~Rovelli, {\it {LQG vertex with finite
  Immirzi parameter}},  {\em Nucl. Phys. B} {\bf 799} (2008) 136--149,
  [\href{http://arxiv.org/abs/0711.0146}{{\tt arXiv:0711.0146}}].

\bibitem{Livine:2024hhc}
E.~R. Livine, {\it {Spinfoam Models for Quantum Gravity: Overview}},
  \href{http://arxiv.org/abs/2403.09364}{{\tt arXiv:2403.09364}}.

\bibitem{Conrady:2010kc}
F.~Conrady and J.~Hnybida, {\it {A spin foam model for general Lorentzian
  4-geometries}},  {\em Class. Quant. Grav.} {\bf 27} (2010) 185011,
  [\href{http://arxiv.org/abs/1002.1959}{{\tt arXiv:1002.1959}}].

\bibitem{Conrady:2010vx}
F.~Conrady, {\it {Spin foams with timelike surfaces}},  {\em Class. Quant.
  Grav.} {\bf 27} (2010) 155014, [\href{http://arxiv.org/abs/1003.5652}{{\tt
  arXiv:1003.5652}}].

\bibitem{Rovelli1995}
C.~Rovelli and L.~Smolin, {\it {Discreteness of area and volume in quantum
  gravity}},  {\em Nuclear Physics B} {\bf 442} (May, 1995) 593--619.

\bibitem{ALarea}
A.~Ashtekar and J.~Lewandowski, {\it {Quantum theory of geometry. 1: Area
  operators}},  {\em Class.Quant.Grav.} {\bf 14} (1997) A55--A82,
  [\href{http://arxiv.org/abs/gr-qc/9602046}{{\tt gr-qc/9602046}}].

\bibitem{Liu:2018gfc}
H.~Liu and M.~Han, {\it {Asymptotic analysis of spin foam amplitude with
  timelike triangles}},  {\em Phys. Rev. D} {\bf 99} (2019), no.~8 084040,
  [\href{http://arxiv.org/abs/1810.09042}{{\tt arXiv:1810.09042}}].

\bibitem{Bianchi:2010fj}
E.~Bianchi, D.~Regoli, and C.~Rovelli, {\it {Face amplitude of spinfoam quantum
  gravity}},  {\em Class. Quant. Grav.} {\bf 27} (2010) 185009,
  [\href{http://arxiv.org/abs/1005.0764}{{\tt arXiv:1005.0764}}].

\bibitem{Han:2013gna}
M.~Han and T.~Krajewski, {\it {Path Integral Representation of Lorentzian
  Spinfoam Model, Asymptotics, and Simplicial Geometries}},  {\em Class. Quant.
  Grav.} {\bf 31} (2014) 015009, [\href{http://arxiv.org/abs/1304.5626}{{\tt
  arXiv:1304.5626}}].

\bibitem{Han:2021rjo}
M.~Han and H.~Liu, {\it {Analytic continuation of spinfoam models}},  {\em
  Phys. Rev. D} {\bf 105} (2022), no.~2 024012,
  [\href{http://arxiv.org/abs/2104.06902}{{\tt arXiv:2104.06902}}].

\bibitem{Simao:2021qno}
J.~D. Sim\~ao and S.~Steinhaus, {\it {Asymptotic analysis of spin-foams with
  timelike faces in a new parametrization}},  {\em Phys. Rev. D} {\bf 104}
  (2021), no.~12 126001, [\href{http://arxiv.org/abs/2106.15635}{{\tt
  arXiv:2106.15635}}].

\bibitem{BJBCrowley_1979}
B.~J.~B. Crowley, {\it Some generalisations of the poisson summation formula},
  {\em Journal of Physics A: Mathematical and General} {\bf 12} (nov, 1979)
  1951.

\bibitem{Barrett:2009gg}
J.~W. Barrett, R.~J. Dowdall, W.~J. Fairbairn, H.~Gomes, and F.~Hellmann, {\it
  {Asymptotic analysis of the EPRL four-simplex amplitude}},  {\em J. Math.
  Phys.} {\bf 50} (2009) 112504, [\href{http://arxiv.org/abs/0902.1170}{{\tt
  arXiv:0902.1170}}].

\bibitem{Kaminski:2017eew}
W.~Kaminski, M.~Kisielowski, and H.~Sahlmann, {\it {Asymptotic analysis of the
  EPRL model with timelike tetrahedra}},  {\em Class. Quant. Grav.} {\bf 35}
  (2018), no.~13 135012, [\href{http://arxiv.org/abs/1705.02862}{{\tt
  arXiv:1705.02862}}].

\bibitem{Tate:2011rm}
K.~Tate and M.~Visser, {\it {Realizability of the Lorentzian (n,1)-Simplex}},
  {\em JHEP} {\bf 01} (2012) 028, [\href{http://arxiv.org/abs/1110.5694}{{\tt
  arXiv:1110.5694}}].

\bibitem{Bonzom:2009hw}
V.~Bonzom, {\it {Spin foam models for quantum gravity from lattice path
  integrals}},  {\em Phys. Rev. D} {\bf 80} (2009) 064028,
  [\href{http://arxiv.org/abs/0905.1501}{{\tt arXiv:0905.1501}}].

\bibitem{Han:2013hna}
M.~Han, {\it {On Spinfoam Models in Large Spin Regime}},  {\em Class. Quant.
  Grav.} {\bf 31} (2014) 015004, [\href{http://arxiv.org/abs/1304.5627}{{\tt
  arXiv:1304.5627}}].

\bibitem{Hellmann:2012kz}
F.~Hellmann and W.~Kaminski, {\it Geometric asymptotics for spin foam lattice
  gauge gravity on arbitrary triangulations},  {\em arXiv preprint
  arXiv:1210.5276} (2012).

\bibitem{Perini:2012nd}
C.~Perini, {\it {Holonomy-flux spinfoam amplitude}},
  \href{http://arxiv.org/abs/1211.4807}{{\tt arXiv:1211.4807}}.

\bibitem{Engle:2020ffj}
J.~Engle, W.~Kaminski, and J.~Oliveira, {\it Addendum to ‘eprl/fk asymptotics
  and the flatness problem’},  {\em Classical and Quantum Gravity} {\bf 38}
  (2021), no.~11 119401.

\bibitem{10.1007/BFb0074195}
A.~Melin and J.~Sj{\"o}strand, {\it Fourier integral operators with
  complex-valued phase functions},  in {\em Fourier Integral Operators and
  Partial Differential Equations} (J.~Chazarain, ed.), (Berlin, Heidelberg),
  pp.~120--223, Springer Berlin Heidelberg, 1975.

\bibitem{Hormander}
L.~Hormander, {\em The Analysis of Linear Partial Differential Operators I},
  ch.~Chapter 7, p.~Theorem 7.7.5.
\newblock Springer-Verlag Berlin, 1983.

\bibitem{Han:2019vpw}
M.~Han and H.~Liu, {\it {Effective Dynamics from Coherent State Path Integral
  of Full Loop Quantum Gravity}},  {\em Phys. Rev. D} {\bf 101} (2020), no.~4
  046003, [\href{http://arxiv.org/abs/1910.03763}{{\tt arXiv:1910.03763}}].

\bibitem{Han:2021cwb}
M.~Han and H.~Liu, {\it {Loop quantum gravity on dynamical lattice and improved
  cosmological effective dynamics with inflaton}},  {\em Phys. Rev. D} {\bf
  104} (2021), no.~2 024011, [\href{http://arxiv.org/abs/2101.07659}{{\tt
  arXiv:2101.07659}}].

\bibitem{Ashtekar:2006wn}
A.~Ashtekar, T.~Pawlowski, and P.~Singh, {\it {Quantum Nature of the Big Bang:
  Improved dynamics}},  {\em Phys. Rev.} {\bf D74} (2006) 084003,
  [\href{http://arxiv.org/abs/gr-qc/0607039}{{\tt gr-qc/0607039}}].

\bibitem{Han:2022rsx}
M.~Han and H.~Liu, {\it {Covariant ${\bar{\mu}}$-scheme effective dynamics,
  mimetic gravity, and non-singular black holes: Applications to spherical
  symmetric quantum gravity and CGHS model}},
  \href{http://arxiv.org/abs/2212.04605}{{\tt arXiv:2212.04605}}.

\bibitem{Ashtekar:2023cod}
A.~Ashtekar, J.~Olmedo, and P.~Singh, {\it {Regular black holes from Loop
  Quantum Gravity}},  \href{http://arxiv.org/abs/2301.01309}{{\tt
  arXiv:2301.01309}}.

\bibitem{Giesel:2023hys}
K.~Giesel, H.~Liu, P.~Singh, and S.~A. Weigl, {\it {Generalized analysis of a
  dust collapse in effective loop quantum gravity: fate of shocks and
  covariance}},  \href{http://arxiv.org/abs/2308.10953}{{\tt
  arXiv:2308.10953}}.

\bibitem{Giesel:2023reg}
K.~Giesel, H.~Liu, P.~Singh, and S.~A. Weigl, {\it {Embedding regular black
  holes into polymerized effective models and mimetic theory}},  {\em to
  appear}.

\end{thebibliography}\endgroup
\end{document}